\documentclass[aps,pre,preprint,onecolumn,citeautoscript,superscriptaddress,eqsecnum]{revtex4-1}  
\usepackage{amsmath,amssymb,bm} 
\usepackage{graphicx}  
\usepackage[tight]{subfigure}    
\usepackage{color} 
\usepackage[papersize={8.5in,11in}]{geometry}
\usepackage[colorlinks=true]{hyperref}  
\hypersetup{
    bookmarks=true,         
    unicode=false,          
    pdftoolbar=true,        
    pdfmenubar=true,        
    pdffitwindow=false,     
    pdfstartview={FitH},    
    pdftitle={My title},    
    pdfauthor={Author},     
    pdfsubject={Subject},   
    pdfcreator={Creator},   
    pdfproducer={Producer}, 
    pdfkeywords={keyword1} {key2} {key3}, 
    pdfnewwindow=true,      
    colorlinks=true,       
    linkcolor=magenta, 
    citecolor=blue,        
    filecolor=magenta,      
    urlcolor=blue           
} 

\usepackage{amsfonts}
\usepackage{upgreek}
\usepackage{slashed}
\usepackage{latexsym}

\newcommand{\beq}{\begin{equation}}
\newcommand{\eeq}{\end{equation}}
\newcommand{\nn}{\nonumber \\}

\begin{document}

\title{Hyperscaling at the spin density wave quantum critical point\\ in two dimensional metals}

\author{Aavishkar A. Patel}

\affiliation{Department of Physics, Harvard University, Cambridge MA 02138, USA}

\author{Philipp Strack}

\affiliation{Department of Physics, Harvard University, Cambridge MA 02138, USA}
\affiliation{Institut f\"ur Theoretische Physik, Universit\"at zu K\"oln, D-50937 Cologne, Germany}

\author{Subir Sachdev}

\affiliation{Department of Physics, Harvard University, Cambridge MA 02138, USA}
\affiliation{Perimeter Institute for Theoretical Physics, Waterloo, Ontario, Canada N2L 2Y5}

\date{\today\\
\vspace{0.6in}}

\begin{abstract}
The hyperscaling property implies that spatially isotropic critical quantum states in $d$ spatial dimensions have
a specific heat which scales with temperature as $T^{d/z}$, and an optical conductivity which scales with frequency as $\omega^{(d-2)/z}$ for $\omega \gg T$,
where $z$ is the dynamic critical exponent. We examine the spin-density-wave critical fixed point of metals in $d=2$ found by Sur and Lee
(Phys. Rev. B {\bf 91}, 125136 (2015)) in an expansion in $\epsilon = 3-d$. We find that the contributions of the ``hot spots'' on the Fermi surface to the optical conductivity and specific heat obey hyperscaling (up to logarithms), 
and agree with the results of the large $N$ analysis of the optical
conductivity by Hartnoll {\em et al.} (Phys. Rev. {\bf 84}, 125115 (2011)). With a small bare velocity of the boson associated with the 
spin density wave order, there is an intermediate energy regime where hyperscaling is violated with $d \rightarrow d_t$, 
where $d_t = 1$ is the number of dimensions transverse to the Fermi surface. 
We also present a Boltzmann equation analysis which indicates that the hot spot contribution to the
DC conductivity has the same scaling as the optical conductivity, with $T$ replacing
$\omega$.
\end{abstract}

\maketitle

\section{Introduction}
\label{sec:intro}

The anomalous properties of the `strange metal' phase of the cuprates, and other correlated electron compounds,
have remained a long-standing challenge to quantum many-body theory. Strange metals are states of quantum matter whose density
can be continuously varied by an external chemical potential at zero temperature, but unlike in a Fermi liquid, there are no long-lived quasiparticle
excitations. It is generally believed that strange metals should be described
by a strongly-coupled quantum-critical theory \cite{SSBK11}, 
but such a proposal immediately faces an obstacle. Almost all strongly-coupled
quantum-critical states, including all conformal field theories, obey the `hyperscaling' property \cite{MEF78}: this implies that the specific
heat, $C_V$, and the conductivity, $\sigma$, scale as
\beq
C_V \sim T^{d/z} \quad; \quad \sigma (\omega \gg T) \sim \omega^{(d-2)/z}, \label{hyper}
\eeq
where $\omega$ is frequency, 
$T$ is temperature, $d$ is the spatial dimension, and $z$ is the dynamic critical exponent; we will refer to the 
conductivity in the $\omega \gg T$ regime above as the optical conductivity.
In the important spatial dimension of $d=2$,
this immediately implies that the optical conductivity should be frequency independent, which contradicts the $\sim \omega^{-0.65}$ behavior
observed in the cuprates \cite{marel03,marel06}.

The scaling arguments can also be naively extended to the DC conductivity, which would then imply that $\sigma (\omega \ll T) \sim T^{(d-2)/z}$.
In $d=2$, this contradicts the widely observed `linear-in-$T$ resistivity', $\sigma \sim T^{-1}$. However, DC transport co-efficients are 
sensitive to constraints from momentum conservation, and so the naive application of hyperscaling to DC transport is often
not valid \cite{HKMS,MBH13,HMPS,Blaise13,Blaise14,Davison14,AAPSS14,Davison15,BD15,ALSS15}. 
But this sensitivity does not extend to the optical conductivity, and so the observations of Ref.~\onlinecite{marel03,marel06}
are the stronger challenge to the hyperscaling property.

There is a much-studied \cite{lee89,Polchinski94,Nayak94a,Nayak94b,lee94,Vadim01,Metzner06,SSL09,MMSS10a,Mross10,Metzner11,SSL13,MMSS14,SSL14,Metzner15} strongly-coupled quantum-critical point which violates hyperscaling: this is the critical point to the onset of
Ising-nematic order in a metal in $d=2$. A closely-related critical theory applies to a $d=2$ metal
coupled to an Abelian or non-Abelian gauge field. We write the properties of the Ising-nematic theory 
in a suggestive form similar to Eq.~(\ref{hyper})
\beq
C_V \sim T^{d_t/z} \quad; \quad \sigma (\omega \gg T) \sim \omega^{(d_t-2)/z}, \quad \quad \mbox{with hyperscaling violation,}
\label{hyperv}
\eeq
where $z=3/2$ is the `fermionic' dynamic critical exponent (in the notation of Ref.~\onlinecite{Mross10}).
For the specific heat, the hyperscaling-violating 
dimensionality $d_t = 1$ has been connected to the number of dimensions transverse to
the Fermi surface \cite{MMSS14,HSS11}. This value of $d_t$ also happens to yield the correct behavior of the optical conductivity in Eq.~(\ref{hyperv}),
although the existing \cite{lee94,HHMS11} physical 
interpretations of this result are different. It is also notable that $\sigma \sim \omega^{-2/3}$ is close
to the experimental observations \cite{marel03,marel06}.

The above violation of hyperscaling is in a theory with a `critical Fermi surface'. On the other hand, theories with Dirac fermions, which are gapless
only at points in the Brillouin zone do obey hyperscaling. 

\begin{figure}
\includegraphics[width=130mm]{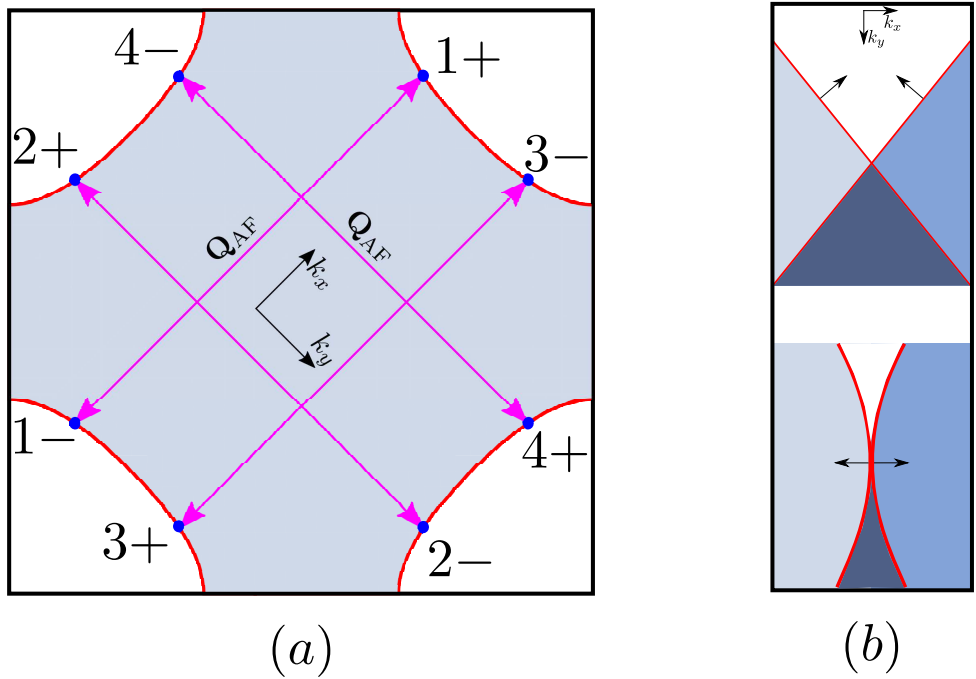}
\caption{
(a) Hot spot geometry, labelling conventions, and choice of $x,y$-coordinate 
system in the Brillouin zone of the two-dimensional square lattice in which 
the fermions move. The boundary of the blue area denotes the Fermi surface separating 
the filled particle-like states (blue area) from the the hole-like states 
(white area). ${\bf Q}_{\rm AF}=(\pi,\pi)$ is the (commensurate) antiferromagnetic ordering 
vector that intersects the Fermi surface at 4 pairs of hot spots.
(b) (Top) Fermi surface patches from a hot spot pair connected by ${\bf Q}_{\rm AF}$ centered at a common origin in momentum space. The two light colored regions are the regions occupied by fermions at the two hot spots of the pair respectively, the dark colored region is occupied by fermions at both hot spots, and the white region is unoccupied. The arrows perpendicular to the Fermi surfaces denote the directions of the Fermi velocities. (Bottom) Under the RG flow, the Fermi surfaces are deformed as shown at the strange metal fixed point, and as indicated in Eq.~(\ref{fslog}). The Fermi velocities are exactly antiparallel only at the hot spot ($\mathbf{k}=0$).
}
\label{fig:hotness}
\end{figure}

Our interest in this paper is the onset of spin density wave order in two-dimensional metals, whose critical theory is described by isolated
points called `hot spots' which are connected to a gapless Fermi surface (see Fig.~\ref{fig:hotness}).
This transition is therefore intermediate between the critical Fermi surface and critical Fermi point cases.
Its field theory \cite{abanov00} 
has a bosonic order parameter $\vec{\phi}$ coupled to fermionic excitations at 
4 pairs of hot spots around the Fermi surface.

In a large $N$ analysis of such a field theory, it was found \cite{abanov00,MMSS10b} that at the two loop level that the Fermi surfaces near the hot spots became asymptotically nested at low energies. In terms of momenta $k_x, k_y$ measuring deviations from the hot spots, the Fermi surface is given by (see the bottom panel of Fig.~\ref{fig:hotness}b)
\beq
k_y \sim \pm \frac{k_x}{\ln (1/|k_x|)}.
\label{fslog}
\eeq
The optical conductivity of the hot spots was computed by Hartnoll {\em et al.} \cite{HHMS11} in a Eliashberg framework, and they
found (at variance with an earlier treatment by Abanov {\em et al.} \cite{ACS03}, and that in Ref.~\onlinecite{CMY14}) a hot spot contribution $\sigma (\omega) \sim \omega^{r_0}$, where the exponent $r_0 > 0$ was determined by the angle
between the Fermi surfaces at the hot spots. For the asympotically nested Fermi surface in Eq.~(\ref{fslog}), it was found \cite{HHMS11}
that $r_0 \rightarrow 0$, indicating that the optical conductivity is a constant (up to logarithms), and so obeys hyperscaling as in 
Eq.~(\ref{hyper}) in $d=2$.

This paper will re-examine these issues using the fixed point for the spin density wave critical found by Sur and Lee \cite{sur15}
using an expansion in $\epsilon = 3-d$. They also also found the asymptotically nested Fermi surfaces in Eq.~(\ref{fslog}) under the 1-loop
renormalization group flow of the $\epsilon$ expansion. We will review their RG analysis in Section~\ref{sec:model}. 
We then proceed to a computation of the optical conductivity in Section~\ref{sec:jj}, and find that the hot
spot contribution obeys the hyperscaling of Eq.~(\ref{hyper}) (up to logarithmic corrections) in the $\epsilon$ expansion, 
in agreement with Hartnoll {\em et al.} \cite{HHMS11}.
We turn to a computation of the non-zero temperature free energy density in the $\epsilon$ expansion in Section~\ref{sec:free}.
We find a result for the hot spot contribution to the specific heat again in agreement with the hyperscaling of Eq.~(\ref{hyper}), and for reasons similar to those for the optical
conductivity.

Sections~\ref{sec:jj} and~\ref{sec:free} also examine the optical conductivity and the free energy in the limit of a
vanishing bare $\vec{\phi}$ velocity: $c \rightarrow 0$. As the bare velocity is generically finite, such a limit can only
apply to observable properties over intermediate $\omega$ or $T$: we find the allowed range is $c \Lambda < \omega, T < v_F \Lambda$, 
where $v_F$ is a Fermi velocity (see Eq.~(\ref{eq:dispersions})), and $\Lambda$ is high momentum cutoff. 
Only in such a limit do we find hyperscaling violation
as described by Eq.~(\ref{hyperv}) with $d_t = 1$. The quantum critical optical conductivity 
studied in Refs.~\onlinecite{ACS03,CMY14} is analogous to this intermediate regime, and we maintain that their results do not
apply when the the bare velocity $c$ is not small.

The more subtle question of the DC conductivity is examined in Section~\ref{sec:boltzmann}; in discussions of the DC conductivity, 
we implicitly assume that $\omega \ll T$. Here, we have to consider the interplay between the hot spots
on the Fermi surface with the remainder of the `cold' Fermi surface more carefully \cite{Rice95,Rosch00,HHMS11,AAPSS14}. 
The cold fermions can short-circuit electronic transport, and so possibly
dominate the DC conductivity. More generally, this belongs to a class of effects associated with the conservation of total momentum,
which can relax only via quenched disorder or umklapp scattering beyond that already continued in the continuum theory \cite{AAPSS14}. 
A general framework for describing such effects was presented
in Refs.~\onlinecite{HKMS,ALSS15}, using solvable holographic models, relativistic hydrodynamics, and memory functions.
In the context of strange metals, it useful to begin with a microscopic model in which total momentum is exactly conserved \cite{HMPS,AAPSS14}.
Then the conductivity can be written as \cite{HKMS}
\beq
\sigma = \sigma_Q + \frac{\mathcal{Q}^2}{\mathcal{M}} \, \frac{1}{(-i \omega)}, \label{sQ}
\eeq
where $\sigma_Q$ is a finite and $T$-dependent `quantum critical' conductivity, and the second term 
can be viewed as the contribution of the cold Fermi surface. The pole at $\omega=0$ has a co-efficient determined 
by static thermodynamic susceptibilities associated with the electric current $J$ and the momentum density $P$, with $\mathcal{Q} = \chi_{JP}$
and $\mathcal{M} = \chi_{PP}$. These thermodynamic susceptibilities are usually non-critical, and so can be taken to be non-universal
and $T$-independent constants, which depend on the full short-distance structure of the theory. Now we add perturbations associated with
umklapp scattering or quenched disorder which can relax the total momentum \cite{HKMS,MBH13,HMPS,Blaise13,Blaise14,Davison14,AAPSS14,Davison15,BD15,ALSS15,AL15,Donos15,Banks15,AL15a,GLSS15}: this leads to a momentum
relaxation rate $\Gamma$ which shifts the pole in Eq.~(\ref{sQ}) off the real axis to $\omega = - i \Gamma$, and so the conductivity takes
the finite value at $\omega=0$
\beq
\sigma = \sigma_Q + \frac{\mathcal{Q}^2}{\mathcal{M}} \, \frac{1}{(-i \omega + \Gamma)}. \label{sQG}
\eeq
Note that $\Gamma$ does have a singular $T$ dependence associated with universal properties of the quantum-critical theory,
and can be computed via memory functions \cite{HKMS,HH12,HST12,HST12a,MBH13,HMPS,LSS14,AAPSS14,AL15}.
A notable feature \cite{DVK15} of Eq.~(\ref{sQG}) is that the 
quantum-critical $\sigma_Q$ and the momentum-mode conductivity are additive; this is in
contrast to the Matthiessen's Rule for quasiparticle theories, in which different quasiparticle scattering mechanisms are additive in the resistivity. 
The $T$ dependence of the momentum-mode term in Eq.~(\ref{sQG}) was discussed in Ref.~\onlinecite{AAPSS14}, using the assumption that
the cold regions of the Fermi surface are `lukewarm' {\em i. e.\/} the electron-electron scattering rate on the entire Fermi surface is faster than the impurity
scattering rate; the results of Ref.~\onlinecite{AAPSS14} are not modified by the analysis of the present paper.

Section~\ref{sec:boltzmann} will present a computation of the quantum-critical conductivity $\sigma_Q$ for the case of a spin density wave quantum critical point in a metal in $d=2$. The momentum mode contribution in Eq.~(\ref{sQG}) was computed in a previous work by two of us \cite{AAPSS14}, and will not be addressed here. 
The computation of $\sigma_Q$ here is aided by the fact that the theory
describing the hot spots is particle-hole symmetric. This implies that the scaling limit theory has $\mathcal{Q}=0$, and so we can
cleanly separate away the momentum mode contribution; the full theory ultimately has $\mathcal{Q} \neq 0$, but this arises from portions of the Fermi surface away from the hot spots \cite{AAPSS14}. Such a separation between $\sigma_Q$ and the momentum mode is more complicated
in general \cite{Davison15}: in particular, for the Ising-nematic critical point there is no particle-hole symmetry to aid us, and we are not aware
of any computation of $\sigma_Q$ for this case. For the spin density wave critical point, we compute $\sigma_Q$
in Section~\ref{sec:boltzmann} using a Boltzmann equation method developed for conformal field theories \cite{damle97,SS98,piazza14,kamenev11}.
We will carry out the Boltzmann analysis directly in $d=2$, rather than the technically more cumbersome $\epsilon$ expansion.
Consequently, our results for $\sigma_Q$ will be qualitative, and not systematic. From the computations in Section~\ref{sec:boltzmann},
we estimate that the leading $T$-dependence of $\sigma_Q$ has the same form as the $\omega$-dependence of the optical conductivity:
{\em i.e.\/} with bare velocities finite, hyperscaling is preserved with $\sigma_Q \sim$ constant; and with vanishing bare velocities, there is 
violation of hyperscaling with $\sigma_Q \sim T^{(d_t-2)/z}$ and $d_t=1$ and over intermediate $T$ range $c \Lambda < T < v_F \Lambda$.

\section{Model}
\label{sec:model}

In this section, we first recapitulate the low-energy continuum quantum field 
theory for fermions moving in a two-dimensional square lattice close to the 
transition to the antiferromagnetic phase with commensurate 
ordering wave vector ${\bf Q}_{\rm AF} = (\pi, \pi)$ \cite{abanov00,MMSS10b,sur15}.
We then explain the embedding by Sur and Lee \cite{sur15} of the 
two-dimensional system into a higher-dimensional $d=3-\epsilon$ space,
and summarize the basic features of the $\epsilon$ expansion.

We begin by defining the action in frequency and momentum representation
$S[\bar{\psi},\psi, \vec{\phi}]$ in two space dimensions $x$ and $y$ 
and one temporal (imaginary time) direction $\tau$:
\begin{align}
S[\bar{\psi},\psi,\vec{\phi}]=
&\sum_{\ell=1}^4 \sum_{m = \pm} \sum_{\sigma=\uparrow,\downarrow}
\int_{k}
\bar{\psi}^{(m)}_{\ell,\sigma}(k) 
\left[ i k_\tau + e^m_n(\mathbf{k})\right]
\psi^{(m)}_{\ell,\sigma}(k) 
+
\frac{1}{2}
\int_q
\vec{\phi}(-q)\cdot
\left[q_\tau^2 + c^2 \mathbf{q}^2 + r\right]
\vec{\phi}(q)
\nonumber\\
&
+g \sum_{\ell=1}^4 \sum_{\sigma,\sigma' = \uparrow,\downarrow}
\int_k \int_q
\left[
\vec{\phi}(q) \cdot \bar{\psi}^{(+)}_{\ell,\sigma}(k+q)
\vec{\tau}_{\sigma,\sigma'}
\psi^{(-)}_{\ell,\sigma'}(k)
+ h.c.
\right].
\label{eq:action}
\end{align}
Here, the functional integral for the fermions goes over fermionic Grassmann fields $\bar{\psi}$, $\psi$ 
which carry additional labels according to their ``home'' hot spot (depicted in Fig.~\ref{fig:hotness}). 
Via a ``Yukawa'' coupling $g$, the fermions are (strongly) coupled to a bosonic vector field with 
three components $\vec{\phi}$  whose fluctuations represent spin waves. At zero temperature $k_\tau$ is a continuous (imaginary)
frequency variable with $k=(k_\tau,\mathbf{k})=(k_\tau,k_x,k_y)$ and likewise for $q$.

According to Fig.~\ref{fig:hotness}, the dispersions of the fermions $e_\ell^\pm(\mathbf{k})=\mathbf{v}_\ell^\pm\cdot\mathbf{k}$ in the hot regions are
\begin{align}
e^{\pm}_1(\mathbf{k}) &= -e^{\pm}_3(\mathbf{k}) = v_F \left( v k_x \pm k_y \right)
\nonumber\\
e^{\pm}_2(\mathbf{k}) &= -e^{\pm}_4(\mathbf{k}) = v_F \left(\mp k_x + v k_y \right)\;,
\label{eq:dispersions}
\end{align}
with $v$ being the ratio of the velocities in $x$ and $y$-direction; we will henceforth set $v_F =1$.  
In particular, the limit $v \rightarrow 0$ corresponds  to locally nested pairs of hot spots, in which the Fermi line becomes orthogonal to the antiferromagnetic ordering vector ${\bf Q}_{\rm AF}$  and the fermion becomes one-dimensional and disperses parallel to ${\bf Q}_{\rm AF}$.

The physics of the action Eq.~(\ref{eq:action}) in two space dimensions has been addressed with a variety of techniques including resummation of subclasses of Feynman 
diagrams \cite{abanov00}, field-theoretic renormalization group techniques \cite{MMSS10b,HHMS11}, and Polchinski-Wetterich flow equations for the effective action \cite{lee13}.
The bottom line is that the fermions and spin-waves are strongly coupled, one has to account for strong renormalization of the shape of the Fermi surface \cite{MMSS10b}. 

Here we embed the fermionic system in two space dimensions described by Eq.~(\ref{eq:action}) into a higher-dimensional space; the ``extra dimensions'' are 
added perpendicular to the physical Fermi surface \cite{sur15} that lies in the $x$-$y$ plane  and has co-dimension 1. Artificially introduced Fermi surfaces with co-dimension $>1$ are gapped out by assuming a $p$-wave charge density wave order in directions perpendicular to the physical Fermi surface. This results in line nodes of the fermionic 
dispersion with co-dimension 1 as needed. The main advantage of this embedding is that the density of states at the Fermi line is suppressed to $\rho(E)\sim E^{d-2}$, that is, it vanishes with energy for $d>2$. This allows the powerful dimensional regularization techniques of relativistic systems to be adapted to the present problem. 

The $d+1$-dimensional action
\begin{align}
S=& \sum_{n=1}^4 \sum_{\sigma = 1}^{N_c} \sum_{j=1}^{N_f}
\int_{k}
\bar{\Psi}_{n,\sigma,j}(k) \left[ i\mathbf{\Gamma} \cdot \mathbf{K} + i \gamma_{d-1} \varepsilon_n(\mathbf{k})\right]
\Psi_{n,\sigma,j}(k)
+
\frac{1}{4}
\int_{q}
\left[|\mathbf{Q}|^2+c^2 \mathbf{q}^2\right]
{\rm Tr}\left[\Phi(-q)\Phi(q)\right]
\nonumber\\
&+
\frac{g\mu^{(3-d)/2}}{\sqrt{N}_f}
\sum_{n=1}^4 \sum_{\sigma,\sigma'=1}^{N_c} \sum_{j=1}^{N_f}
\int_{k}\int_{q}
\left[\bar{\Psi}_{n,\sigma,j}(k+q) 
\Phi_{\sigma,\sigma'}(q) i \gamma_{d-1} 
\Psi_{\bar{n},\sigma',j}(k)+h.c.\right] 
\label{eq:embedded_action}
\end{align}
is integrated over $k=(\mathbf{K},\mathbf{k})$, which contains the physical momentum $\mathbf{k}=(k_x,k_y)$ 
and a $d-1 = 2-\epsilon$ dimensional ``generalized frequency'' vector $\mathbf{K} = (k_\tau,\mathbf{\bar{K}})=(k_\tau, k_1,...,k_{d-2})$, that includes the physical frequency 
$k_\tau$ in its first component and the $d-2$ extra dimensions in the others and likewise for $q$. The bosons have been promoted to matrix fields $\phi(q) = \sum_{a=1}^{N_c^2 - 1}
\phi^a(q) \tau^a$ with Tr$\left[\tau^a\tau^b\right] = 2 \delta^{ab}$ conventions for the trace over SU$(N_c)$ generators $\tau^a$. The fermions are collected in a SU($N_f)$ 
flavor group and the physical limit of Eq.~(\ref{eq:embedded_action}) is
\begin{align}
\mathbf{K}\rightarrow k_\tau\;,\;\;\;\epsilon \rightarrow 1\;,\;\;\;\; d\rightarrow2\;,\;\;\;N_c = 2\;,\;\;\;N_f = 1.
\label{eq:physical}
\end{align}
Computations with Eq.~(\ref{eq:embedded_action}) involve traces over products of $d-1$ dimensional gamma matrices, collected in the vector $(\mathbf{\Gamma},\gamma_{d-1})$ with 
$\mathbf{\Gamma} = (\gamma_0,\mathbf{\bar{\Gamma}}) = (\gamma_0,\gamma_1,...,\gamma_{d-2})$, that satisfy $\left\{\gamma_\mu,\gamma_\nu\right\} = 2 I \delta_{\mu\nu}$ and Tr~$I=2$. The book-keeping indices for the hot spots are: $\bar{1}=3$, $\bar{2}=4$, $\bar{3}=1$, $\bar{4}=2$; the two-component fermion spinors appearing in Eq.~(\ref{eq:embedded_action}) disperse according to
\begin{align}
\varepsilon_1(\mathbf{k})& = e^+_1(\mathbf{k}),
\nonumber\\
\varepsilon_2(\mathbf{k})& = e^+_2(\mathbf{k}),
\nonumber\\
\varepsilon_3(\mathbf{k})& = e^-_1(\mathbf{k}),
\nonumber\\
\varepsilon_4(\mathbf{k})& = e^-_2(\mathbf{k}),
\label{ee1}
\end{align}
with the right-hand-sides defined in Eq.~(\ref{eq:dispersions}). The two-component spinors of Eq.~(\ref{eq:embedded_action}) 
contain two of the original fermions from opposing sides of the Fermi surface \cite{sur15}.

Sur and Lee \cite{sur15} performed a field-theoretic one-loop renormalization group analysis of Eq.~(\ref{eq:embedded_action}) in $d=3-\epsilon$ dimensions. They retained 
the simplest set of 5 independent running couplings. For the fermion propagator 3  wave-function renormalization factors are used, one in the direction of  ``time and extra dimensions'' $\mathbf{K}$ and one each in  the $k_x$  and $k_y$ directions. For the Bose propagator there are 2 wave-function renormalization factors, one in the $\mathbf{Q}$ 
direction and one for the $q_{x,y}$ directions (which have to be equivalent by point group symmetry). 

The fixed point of the $\epsilon$ expansion is defined in terms of the ratios $\lambda= g^2/v$ and $w=v/c$: 
\begin{align}
&\lambda \rightarrow \lambda^\ast = 4\pi \epsilon \frac{N_c^2 + N_c N_f -1}{N_c^2 + N_c N_f -3}, \nonumber \\
&w \rightarrow w^\ast = \frac{N_c N_f}{N_c^2-1},
\end{align}
The fixed point determines a dynamic critical exponent $z$ via 
\begin{align}
z= 1+ \frac{\lambda^\ast}{8\pi} + \mathcal{O}(\epsilon^2).
\label{eq:z}
\end{align}
Note that $k_\tau$ and $\mathbf{\bar{K}}$ scale as $k_y^z$; so 
the extra $d-3$ spatial dimensions and the time dimension \textit{both} scale as $z$ with respect to the two physical dimensions, 
instead of just the time dimension as is usually the case with other models.

While the scaling structure described so far is conventional, there are logarithmic corrections which arise
from the flow of the velocities $v$ and $c$ flow to zero at long length scales. This flow is described by 
\begin{align}
&\frac{dc}{d\ln \mu}\approx\frac{4z}{\pi}(z-1)c^2, \nonumber \\
&\frac{dv}{d\ln \mu}\approx w^\ast \frac{dc}{d\ln \mu},
\label{eq:vcflow}
\end{align}
where $\mu$ is the renormalization group momentum scale. Such a dynamic nesting with $v \rightarrow 0$
was found also in an earlier $1/N_f$ expansion \cite{MMSS10b}. At the fixed point with vanishing $v$ and $c$, the antiferromagnetic ordering vectors intersect the Fermi surface at a right angle. This is illustrated in Fig.~\ref{fig:hotness}(b).
Note that with $v \rightarrow 0$ at the fixed point, we must also have $g^2 \rightarrow 0$ for the coupling $\lambda$ to remain finite; this is indeed found to be the case in the renormalization group flow.

\section{Optical conductivity $\sigma (\omega)$}
\label{sec:jj}

In this section, we compute the optical conductivity $\sigma(\omega)$ for fermions near the hot spots at the $\epsilon$ expansion
fixed point described in Section~\ref{sec:model}. Our computation will be to order $\epsilon$, which requires evaluation of two-loop Feynman
graphs. 
 
Before embarking on the description of the Feynman graphs, let us review the expectations of a general scaling analysis.
The spatial directions, $x$, $y$, have scaling dimension 1, the time direction has scaling dimension $z$, and the $1-\epsilon$ extra spatial directions with Dirac dispersion also have scaling dimension $z$. So the scaling dimension of the free energy density, $F$, is
\beq
[F] = 2 + (2-\epsilon)z. \label{Fdim}
\eeq
The vector potential, $A$, has dimension 1, and so the electric current, $J$, being proportional to $\delta F/\delta A$ has dimension
\beq
[J]  = 1 + (2-\epsilon) z. \label{Jdim}
\eeq
Finally, the conductivity is given by the Kubo formula in terms of a current correlator, from which we deduce
\beq
[\sigma] = (1-\epsilon) z. \label{sdim}
\eeq
These are the scaling expectations for a theory that obeys hyperscaling. 
If we have a violation of hyperscaling, we expect the spatial direction along the Fermi surface to not contribute in the counting of scaling dimension.
So we should have 
\beq
[F] = 1 + (2-\epsilon)z \quad, \quad [\sigma] = -1 + (1-\epsilon)z, \quad \quad \mbox{with hyperscaling violation.} \label{Fsdimv}
\eeq
We already know that the Fermi liquid contribution of the quasiparticles far from the hot spots violates hyperscaling as
in Eq.~(\ref{Fsdimv}) with $z=1$. The question before us is whether the hot spot contribution preserves hyperscaling as in Eqs.~(\ref{Fdim}) 
and (\ref{sdim}), or violates hyperscaling as in Eq.~(\ref{Fsdimv}).
 
We first compute the one-loop (free fermion) contribution to the two-point correlator of the current density $\langle J_y J_y \rangle$. Then, we compute the two-loop (interaction) contributions to $\langle J_y J_y \rangle$, of which there are two: the ``self-energy correction'' (Section~\ref{subsec:self}) and the ``vertex correction'' (Section~\ref{subsec:vert}).
Finally, in Section~\ref{subsec:kubo} we compile the results from the evaluation of the loop diagrams and, applying the Kubo formula, we derive the scaling form of the optical conductivity $\sigma (\omega)$ for the fermions near the hot spots. 

\subsection{One-loop contribution to $\langle J_y J_y \rangle$}
\label{subsec:one-loop}
We have, for the current density in the $y$ direction,
\begin{align}
&J_y = J_y^{(1)} + J_y^{(3)}+J_y^{(2)} + J_y^{(4)} =  \nonumber \\ 
&i\sum_{\sigma=1}^{N_c}\sum_{j=1}^{N_f}(\bar{\Psi}_{1,\sigma,j}\gamma_{d-1}\Psi_{1,\sigma,j} - \bar{\Psi}_{3,\sigma,j}\gamma_{d-1}\Psi_{3,\sigma,j})+iv^2\sum_{\sigma=1}^{N_c}\sum_{j=1}^{N_f}(\bar{\Psi}_{2,\sigma,j}\gamma_{d-1}\Psi_{4,\sigma,j} + \bar{\Psi}_{2,\sigma,j}\gamma_{d-1}\Psi_{4,\sigma,j}),
\label{eq:current_operator}
\end{align}
and likewise for $J_x$ but with $(1,3)\leftrightarrow(4,2)$.

\begin{figure}
\includegraphics[width=120mm]{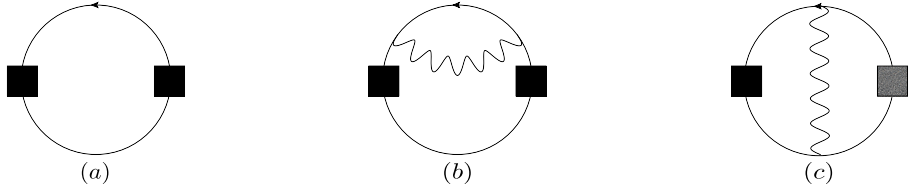}
\caption{Feynman graphs for the current-current correlator up to two loops. The black and grey boxes 
are current vertices for the $n$ and $\bar{n}$ hot spot pairs respectively. The wiggly lines are the boson propagators and the solid lines stand for 
fermion propagators. (a): One-loop contribution for free fermions. (b): Two-loop self-energy 
correction computed in Section~\ref{subsec:self}. There is also a partner diagram with the boson on the lower fermion line. (c): Two-loop vertex correction 
computed in Section~\ref{subsec:vert}.} 
\label{fig:JJ_graphs}
\end{figure}

The one-loop contribution to this correlator is simply the non-interacting ``bubble'' containing 
a convolution of two fermion propagators as shown in Fig.~\ref{fig:JJ_graphs}, both from the {\it same} hot spot (we follow the 
index convention of Eq.~(\ref{ee1}) and absorb the identical contributions from the other hot spots into a prefactor)
\begin{align}
\langle J_yJ_y \rangle_{\rm{1-loop}}(\omega)=-2(1+v^2) N_cN_f 
\int\frac{d^2\mathbf{k}}{(2\pi)^2}\frac{d^{2-\epsilon}\mathbf{K}}{(2\pi)^{2-\epsilon}}\mathrm{Tr}
\left[i\gamma_{d-1}
G_1(\mathbf{K},\mathbf{k})
i\gamma_{d-1}
G_1(\mathbf{K}+\mathbf{W},\mathbf{k})
\right],
\nonumber\\
\label{eq:bubble}
\end{align}
where $\mathbf{W}=(\omega,\mathbf{\bar{0}})$ and the fermion propagator is given by
\begin{align}
G_n(\mathbf{K},\mathbf{k})=
(-i)\frac{\mathbf{\Gamma}\cdot\mathbf{K} +\gamma_{d-1}\varepsilon_n(\mathbf{k})}{\mathbf{K}^2+\varepsilon_n(\mathbf{k})^2}.
\end{align}
We evaluate Eq.~(\ref{eq:bubble}) using Feynman parameters in App.~\ref{app:one-loop} and 
obtain in Eq.~(\ref{eq:jyjyfree}) to leading order in $\epsilon$:
\begin{align}
&\langle J_yJ_y \rangle_{\rm{1-loop}}(\omega)=
-\sqrt{1+v^2} \int dk_\parallel~N_cN_f\omega^{1-\epsilon} \left(\frac{1}{16\pi} \right).
\label{eq:one-loop}
\end{align}
where $k_\parallel$ is the component of $\mathbf{k}$ along the Fermi surface of $\varepsilon_1 (\mathbf{k})$ (note $k_\parallel=k_x$ for $v=0$).
For comparison with the subsequent two-loop contribution, it is useful to write this as 
\begin{equation}
\langle J_y J_y \rangle_{\rm 1-loop}(\omega)=-(1+v^2) \int \frac{d\varepsilon_3}{2v}N_cN_f\omega^{1-\epsilon}\left(\frac{1}{16\pi}\right),
\label{eq:one-loop2} 
\end{equation}
where the variable of integration $\varepsilon_3$ is a co-ordinate orthogonal to the equal energy lines of $\varepsilon_3 (\mathbf{k})$.

We can evaluate the integral over $\varepsilon_3$ to yield a factor of $\Lambda$, a large-momentum cutoff,
and then we conclude that $\sigma_{\rm{1-loop}} (\omega) \sim \omega^{-\epsilon}$. We now observe that this result
agrees with hyperscaling violating scaling dimension in Eq.~(\ref{Fsdimv}) for $z=1$. This is just the expected result, because
we are dealing with the contribution of free fermions, and there is no distinction yet between the hot-spot contribution, and the 
Fermi liquid contribution of quasiparticles far from the hot spot.

\subsection{Two-loop self-energy correction $\langle J_y J_y \rangle_{\rm SE}$}
\label{subsec:self}
\begin{figure}[t]
\includegraphics[width=85mm]{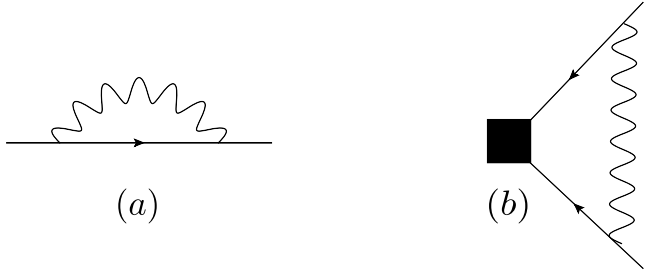}
\caption{Key one-loop elements appearing in the two-loop self-energy correction 
(a) and two-loop (current) vertex correction (b).} 
\label{fig:1-loop}
\end{figure}

To investigate the impact of interactions on $\langle J_y J_y \rangle$, we first compute the two-loop self-energy correction 
depicted in Fig.~\ref{fig:JJ_graphs}(b). There are two diagrams here with identical contributions, whose sum gives
\begin{align}
\langle J_y J_y \rangle_{\mathrm{SE}}(\omega)=-4(1+v^2)N_f \int \frac{d^2\mathbf{k}}{(2\pi)^2}\frac{d^{2-\epsilon}\mathbf{K}}{(2\pi)^{2-\epsilon}} \mathrm{Tr}
\Bigg[&i\gamma_{d-1}
G_1(\mathbf{K},\mathbf{k}) 
\Sigma_1(\mathbf{K},\mathbf{k})
G_1(\mathbf{K},\mathbf{k}) 
i\gamma_{d-1}
G_1(\mathbf{K}+\mathbf{W},\mathbf{k})
\Bigg].
\label{eq:two-loop-self}
\end{align}
Here we note that this expression contains three fermion propagators from the same 
hot spot pair $1$ and one, inside the one-loop self-energy $\Sigma_n(\mathbf{K},\mathbf{k})$, 
depicted in Fig.~\ref{fig:1-loop} (a) from its ``partner'' hot spot pair $3$.
We now compute $\Sigma_1(\mathbf{K},\mathbf{k})$ separately. After that, we substitute the 
result back into Eq.~(\ref{eq:two-loop-self}) and perform the remaining integrations over $\mathbf{k}$ and $\mathbf{K}$.
The effect of large momentum-transfer scattering of fermions from one hot spot pair ($1$)
to its partner ($3$) via exchange of bosonic spin fluctuations is captured in the self-energy
\begin{align}
\Sigma_1(\mathbf{K},\mathbf{k})= \frac{g^2\mu^\epsilon}{N_f}\sum_{j=1}^{N_c^2-1}(\tau^j\tau^j)
\int \frac{d^2\mathbf{q}}{(2\pi)^2}\frac{d^{2-\epsilon}\mathbf{Q}}{(2\pi)^{2-\epsilon}}
i\gamma_{d-1}
G_{3}(\mathbf{Q}+\mathbf{K},\mathbf{q}+\mathbf{k})
i\gamma_{d-1}
D(\mathbf{Q},\mathbf{q})\;,
\label{eq:oneloop_self}
\end{align}
where the spin fluctuation propagator
\begin{align}
D(\mathbf{Q},\mathbf{q}) = \frac{1}{\mathbf{Q}^2+c^2 \mathbf{q}^2} 
\end{align}
involves the spin-wave velocity $c$ which vanishes at the Sur-Lee 
fixed point near the hot spots, as does the Yukawa coupling $g^2$; the ratio $g^2/c$ however 
attains a finite value (see Sec.~\ref{sec:model}).

We evaluate the expressions first in Appendix~\ref{app:self-one-loop} using a simplifying approximation valid only for small bare
velocities $c$ and $v$. In the limit $v,c \rightarrow 0$, the integrand in Eq.~(\ref{eq:oneloop_self}) then depends on $q_x$ only via the spin fluctuation propagator. Thus, we can first 
perform the $q_x$ integration and then set $c=0$, which is equivalent to replacing 
\begin{align}
\int \frac{d^2 \mathbf{q}}{(2 \pi)^2}  & \rightarrow \int \frac{d q_y}{(2 \pi)^2}
\nonumber\\
D(\mathbf{Q},\mathbf{q}) &\rightarrow \frac{\pi}{c} \frac{1}{|\mathbf{Q}|}\;.
\label{eq:quasilocaltrick}
\end{align}
in Eq.~(\ref{eq:oneloop_self}).
This way, Eq.~(\ref{eq:oneloop_self}) picks up the finite prefactor $g^2/c$ 
and the integrand becomes independent of both velocities $v$ and $c$.
The resulting integrals are performed in Appendix~\ref{app:self-one-loop} using Feynman parameters
to obtain
\begin{equation}
\Sigma_1(\mathbf{K},\mathbf{k})= 
\Sigma_1(\mathbf{K})= 
-i\frac{\pi^{2-\epsilon/2}\Gamma(\epsilon/2)}{(2\pi)^{4-\epsilon}}
\frac{\mu^\epsilon}{N_f}\frac{g^2}{c}\sum_{j=1}^{N_c^2-1}(\tau^j\tau^j)
\frac{\mathbf{\Gamma}\cdot\mathbf{K}}
{[\mathbf{K}^2]^{\epsilon/2}}
\int_0^1 dx 
\frac{(1-x)^{\frac{1}{2} - \frac{\epsilon}{2}}}{x^{\frac{1}{2}+\frac{\epsilon}{2}}}\;.
\label{eq:one-loop-self-freq}
\end{equation}
We observe that the above fermion self-energy depends only on frequency and not an spatial momenta $\mathbf{k}$ anymore; 
the fermions near the hot spots see an essentially ``local'' boson with the ``1 over frequency'' propagator of Eq.~(\ref{eq:quasilocaltrick}).  
Eq.~(\ref{eq:one-loop-self-freq}) induces an anomalous scaling for the $\mathbf{K}$ dependent part of the fermion propagator, 
but the absence of anomalous dimensions for the spatial $\mathbf{k}$ components renormalizes the dynamical exponent $z$ to values larger than one at the Sur-Lee fixed point.

However, for our purposes here, the approximation associated with Eq.~(\ref{eq:quasilocaltrick}) turns out not to be sufficient, since $v$ and $c$ vanish only logarithmically near the hot spot.  It is thus crucial to obtain the full $v$ and $c$ dependence of the pole term in Eq.~(\ref{eq:one-loop-self-freq}), and of that in Eq.~(\ref{eq:two-loop-self}). The needed integrals are computed in Appendix~\ref{app:self-one-loop-vcn0}, and the final result  for the two-loop self-energy correction is the rather complicated expression in Eqs.~(\ref{eq:complicated1}) and (\ref{eq:complicated2}). Computing its singular pole in $\epsilon$ and dropping power divergent terms, we obtain
\begin{align}
\langle J_y J_y \rangle_{\mathrm{SE}}(\omega)\approx
\int \frac{d \varepsilon_3}{2 v} ~ \frac{(N_c^2-1)g^2\mu^\epsilon\omega^{1-\epsilon}}{64\pi^3c\epsilon}
\int_0^1 dx \frac{(1-x)^{1/2}(1+v^2)}{(c^2+x(1+v^2-c^2))^{1/2}}\left(\omega^2+\frac{c^2\varepsilon_3^2}{c^2+x(1+v^2-c^2)}\right)^{-\epsilon/2},
\label{eq:self_singular}
\end{align}
in terms of the same $\varepsilon_3$ variable of integration used in Eq.~(\ref{eq:one-loop}).

\subsection{Two-loop vertex correction to $\langle J_y J_y \rangle_{\rm vert}$}
\label{subsec:vert}
The vertex correction graph, Fig.~\ref{fig:JJ_graphs}(c), is considerably more involved than the self-energy correction of the preceding section; although for $c\rightarrow0$
it is free of $1/\epsilon$ poles of the type Eq.~(\ref{eq:self_singular}). Using the abbreviation $\Xi_1(\mathbf{K},\mathbf{k},\mathbf{W})$ for the one-loop current 
vertex correction in Fig.~\ref{fig:1-loop}(b), we can write the entire graph including contributions from all hot spot pairs as
\begin{align}
\langle J_y J_y \rangle_{\mathrm{vert}}(\omega) = -2(1-v^2)iN_f \int\frac{d^2\mathbf{k}}{(2\pi)^2}\int\frac{d^{2-\epsilon}\mathbf{K}}{(2\pi)^{2-\epsilon}}\mathrm{Tr}\Bigg[\gamma_{d-1}G_1(\mathbf{K},\mathbf{k})\Xi_3(\mathbf{K},\mathbf{k},\mathbf{W})G_1(\mathbf{K}+\mathbf{W},\mathbf{k})\Bigg],
\nonumber\\
\label{eq:two-loop-vert}
\end{align}
We observe that Eq.~(\ref{eq:two-loop-vert}) contains two fermion propagators from one hot spot pair and two fermion propagators (inside the current vertex $\Xi_1$) from the partner hot spot pair, unlike the self energy correction Eq.~(\ref{eq:two-loop-self}). The one-loop correction to the $J_y$ vertex,
\begin{align}
&\Xi_3(\mathbf{K},\mathbf{k},\mathbf{W})=\nonumber \\
&i\frac{g^2\mu^\epsilon}{N_f}\sum_{j=1}^{N_c^2-1}(\tau^j\tau^j)\int \frac{d^2\mathbf{q}}{(2\pi)^2}\frac{d^{2-\epsilon}\mathbf{Q}}{(2\pi)^{2-\epsilon}}
\Bigg[
\gamma_{d-1}G_3(\mathbf{K}+\mathbf{Q},\mathbf{k}+\mathbf{q})\gamma_{d-1}G_3(\mathbf{K}+\mathbf{Q}+\mathbf{W},\mathbf{k}+\mathbf{q}) \gamma_{d-1}\frac{1}{\mathbf{Q}^2+c^2 \mathbf{q}^2} 
\Bigg],
\label{eq:1-loop-current}
\end{align}
does not contain a $1/\epsilon$ pole in the limit of $c\rightarrow 0$. This subsequently leads to the lack of a pole in the two-loop vertex correction  (the details of the computation are presented in Appendix~\ref{app:vert-one-loop}). For $v,c\neq0$ however, as described in Appendix~\ref{app:self-one-loop-vcn0}, the vertex correction picks up a small pole with a coefficient of $O(c)$:
\begin{align}
&\langle J_y J_y \rangle_{\mathrm{vert}}(\omega) \approx - \int\frac{d\varepsilon_3}{2v}\frac{(N_c^2-1) g^2c\mu^\epsilon\omega^{1-\epsilon}}{32\pi^3\epsilon}\int_0^1 dx \frac{(1-x)^{1/2}(1-v^2)}{(c^2+x(1+v^2-c^2))^{3/2}} \left(\omega^2+\frac{c^2\varepsilon_3^2}{c^2+x(1+v^2-c^2)}\right)^{-\epsilon/2}.
\label{eq:vert_singular}
\end{align}
However, this is subdominant to the self energy correction in Section~\ref{subsec:self}, as the latter is finite in the limit $c \rightarrow 0$; this is a consequence of a Ward identity discussed in Appendix~\ref{app:self-one-loop-vcn0}.

\subsection{Renormalized conductivity $\sigma (\omega)$}
\label{subsec:kubo}

We can now add the leading free contribution in Eq.~(\ref{eq:one-loop2}), the singular self-energy correction in Eq.~(\ref{eq:self_singular}) and the appropriate counter term
to obtain the renormalization of the conductivity $\sigma (\omega)$ near the fixed point
\begin{align}
&\sigma_{yy}(\omega)\approx (1+v^2) \int \frac{d\varepsilon_3}{2v} \frac{N_c N_f}{16\pi}\omega^{-\epsilon}\int_0^1 dx\Biggl[1+ \nonumber \\
&\frac{(z-1)}{\pi} \frac{(1-x)^{1/2}}{(c^2+x(1+v^2-c^2))^{1/2}}\ln\left(\frac{\omega^2}{\mu^2}+\frac{c^2\varepsilon_3^2/\mu^2}{c^2+x(1+v^2-c^2)}\right)\Biggr]. 
\label{eq:syyfinal}
\end{align}
The interpretation of this central result requires some care in the limit of small $v$ and $c$, and we consider various cases separately below.
The important point here is that the argument of the logarithm is of order $(\omega^2 + (c \varepsilon_3)^2)/\mu^2$, and so the 
renormalization-group-improved perturbation expansion will lead to powers of $(\omega^2 + (c \varepsilon_3)^2)/\mu^2$, in contrast to the power
of $\omega$ alone outside the square bracket. This difference arises because, at leading order, the singular contribution of the hot spot is the same
as the rest of the Fermi surface, while at higher orders there is quasiparticle breakdown only close to the hot spot.
Consequently, there is a modification in 
the nature of the $\varepsilon_3$ integral, where we recall that $\varepsilon_3$ measures distance away from the hot spot along the Fermi surface.

First, let us assume the bare value of $c$ is so small that the $\varepsilon_3$ dependence of the argument of logarithm can be ignored;
this was, effectively, the limit that was implicitly taken in by Abanov {\em et al.} \cite{ACS03}.
This requires that $c \Lambda < \omega$, where $\Lambda$ is the momentum space cutoff. Then, we can easily perform the integral over $x$,
and after re-exponentiating the logarithm in the $\epsilon$ expansion, we conclude that 
\begin{eqnarray}
\sigma_{yy} (\omega) &\sim&  \int \frac{d\varepsilon_3}{2v} \omega^{- \epsilon} \Bigl[ 1 + (z-1) \ln (\omega/\mu) \Bigr] \nn 
&\sim& \Lambda \mu^{1-z} \, \omega^{-\epsilon + (z-1)}. \label{syyhyperv}
\end{eqnarray}
This is the answer expected from the hyperscaling violation case in Eq.~(\ref{Fsdimv}) at this order in $\epsilon$; Abanov {\em et al.} \cite{ACS03} found
$\sigma (\omega ) \sim 1/\sqrt{\omega}$, which is consistent with Eq.~(\ref{Fsdimv}) for their dynamic critical exponent $z=2$. 
Thus the $\omega$ dependence of $\sigma$ violates hyperscaling as in Eq.~(\ref{hyperv}) with $d_t=1$ for $ c \Lambda < \omega < \Lambda$, 
which can be within
a universal regime only if the bare value of $c$ is small enough, as claimed in Section~\ref{sec:intro}.

Next, we consider the more generic case where the bare value of $c$ is of order unity. Then, we can divide the integration over $\varepsilon_3$ 
in Eq.~(\ref{eq:syyfinal}) into two regimes. There is the far from the hot-spot regime where $c \varepsilon_3 \gg \omega$, and the close to 
the hot-spot regime of $c \varepsilon_3 \ll \omega$. The contribution of the close to the hot-spot regime 
is similar to that in Eq.~(\ref{syyhyperv}), except that the upper bound on the integration over $\varepsilon_3$ involves $\omega$
\beq
\sigma_{yy} (\omega) \sim \int_0^{\omega/c} \frac{d\varepsilon_3}{2v} \omega^{-\epsilon + (z-1)}. \label{syyhyper1}
\eeq
Actually, by scaling, we expect the upper bound on the momentum integral over $\varepsilon_3/v$ to scale as $\omega^{1/z}$ at higher
order in $\epsilon$; using such an upper bound in Eq.~(\ref{syyhyper1}) we obtain the generic hot-spot contribution
\beq
\sigma_{yy} (\omega) \sim \omega^{-\epsilon + (z-1) + 1/z}. \label{syyhyper2}
\eeq
To this order in $\epsilon$, this is the scaling expected by the hyperscaling preserving scaling dimension in Eq.~(\ref{sdim}). In comparison
to the hyperscaling violating answer obtained in the direct $v, c\rightarrow 0$ limit used for Eq.~(\ref{syyhyperv}), the conductivity
has acquired an extra factor of $\omega^{1/z}$. So we reach one of our main conclusions, that the hot-spot contribution to the 
conductivity generically obeys hyperscaling as in Eq.~(\ref{hyper}). We have not written out explicit factors 
of $v$ and $c$ in the final scaling forms,
but these are ultimately only expected to yield powers of $(\ln (1/\omega))^{-1}$, and so hyperscaling is only obeyed up to powers
of $\ln (1/\omega)$.

Finally, we also have to consider the contribution of the far from the hot spot regime $c \varepsilon_3 \ll \omega$. In this regime, the
term inside the square brackets in Eq.~(\ref{eq:syyfinal}) is $\omega$-independent, and so we obtain an additional contribution $\sigma \sim \omega^{-\epsilon}$. This is just the additive Fermi liquid contribution of long-lived quasiparticles far from the hot spot.

\section{$T>0$ free energy}
\label{sec:free}

In order to study the finite temperature dynamics of this model, we need to compute the free energy density at $T>0$. The free energy density has contributions from the free fermions, the free bosons, and a ``self energy" correction due to their interactions. 
Following the lessons learned in the analysis of the optical conductivity in Section~\ref{sec:jj}, we will perform the computation here
only in the simpler limit of vanishing velocites $v, c \rightarrow 0$, 
where we can replace the boson propagator by the momentum-independent form
in Eq.~(\ref{eq:quasilocaltrick}). However, as described in Section~\ref{subsec:kubo}, we will assume that the low $T$ hot spot contribution for
the case of finite velocities can be estimated by limiting the range of the fermionic $k_x$ integral (along the Fermi surface) to an upper limit
$\sim T^{1/z}$; here we have assumed the the upper cutoff is determined by $T$ rather than $\omega$ for the optical conductivity
in Section~\ref{subsec:kubo}.

The free fermion, $F_f^0$, and free boson, $F_b^0$ contributions to the free energy density, $F$, 
are obtained straightforwardly to leading order in $\epsilon$ (the prefactor of $4$ in the fermion contribution comes from having $4$ pairs of hot spots) 
\begin{eqnarray}
F_{f}^{0}&=&4N_cN_fT \int \frac{dk_x}{2\pi} \int \frac{dk_yd^{1-\epsilon}\bar{\mathbf{K}}}{(2\pi)^{2-\epsilon}} \ln\left[(1+e^{(k_y^2+\bar{\mathbf{K}}^2)^{1/2}/T})(1+e^{-(k_y^2+\bar{\mathbf{K}}^2)^{1/2}/T})\right] \nonumber \\ 
&=& \int dk_x~N_cN_f T^{3-\epsilon} \left(\frac{3\zeta(3)}{2\pi^2}\right),
\end{eqnarray}
where the infinite temperature-independent constant part was dropped. For the bosons
\begin{equation}
F_{b}^{0}=(1-N_c^2)T \int \frac{d^2\mathbf{q}}{(2\pi)^2} \int \frac{d^{1-\epsilon}\bar{\mathbf{Q}}}{(2\pi)^{1-\epsilon}} \ln\left[1-e^{-(c^2\mathbf{q}^2+\bar{\mathbf{Q}}^2)^{1/2}/T}\right] =\frac{\pi^2}{90c^2}(N_c^2-1)T^{4-\epsilon}.
\end{equation}

\begin{figure}[h]
\includegraphics[width=30mm]{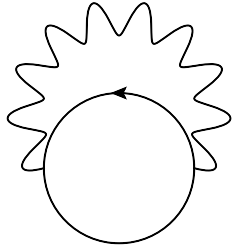}
\caption{The simplest interaction contribution to the free energy at $O(g^2)$.} 
\label{fig:free-energy}
\end{figure}

The interaction contribution to the free energy at two-loop order is given by Fig.~\ref{fig:free-energy}. It may be expressed as
\begin{equation}
F_{fb}=\frac{1}{2}\mathrm{Tr}\left[\sum_{j=1}^{N_c^2-1}\tau^j\tau^j\right]\int\frac{d^2\mathbf{q}d^{1-\epsilon}\mathbf{\bar{Q}}}{(2\pi)^{3-\epsilon}}T\sum_{\omega_q}\frac{\Pi(q,T)}{\mathbf{\bar{Q}}^2+c^2|\mathbf{q}|^2+\omega_q^2},
\end{equation}
where $\Pi(q,T)$ is the RPA polarization bubble given by
\begin{equation}
\Pi(q,T)=g^2\sum_{n}\int\frac{d^2\mathbf{k}d^{1-\epsilon}\bar{\mathbf{K}}}{(2\pi)^{3-\epsilon}}T\sum_{\omega_k}\mathrm{Tr}\left[\gamma_{d-1}G_{n}(k)\gamma_{d-1}G_{n}(k+q)\right].
\end{equation}
We separate out $\Pi(q,T)$ as
\begin{equation}
\Pi(q,T)=\left(\Pi(q,T)-\Pi(q,0)\right)+\Pi(q,0),
\end{equation}
and evaluate the finite temperature part setting $v=0$ at the outset, taking $g,c\rightarrow0$ with $g^2/c$ finite and equal to its fixed point value $\lambda^\ast w^\ast$. The zero temperature part is evaluated with $v\neq0$ at the outset, and $g,v\rightarrow0$ with $g^2/v$ finite and equal to its fixed point value $\lambda^\ast$ . As described in Appendix \ref{app:free-energy}, this separates out the contributions that renormalize the free fermionic and bosonic contributions, with the finite temperature part of $\Pi$ renormalizing the free fermionic contribution and the zero temperature part renormalizing the bosonic contribution. We then obtain (see Appendix~\ref{app:free-energy} for details), for the singular parts,
\begin{align}
&F_{fb}=F_{fb}^{(1)}+F_{fb}^{(2)}, \nonumber \\
&F_{fb}^{(1)}= \int dk_x~(N_c^2-1)T^{3-2\epsilon}\frac{g^2}{c}\left(\frac{3\zeta(3)}{16\pi^3\epsilon}\right), \nonumber \\
&F_{fb}^{(2)}=(N_c^2-1)T^{4-2\epsilon}\frac{g^2\pi}{360v c^2\epsilon}.
\end{align}
We have set the momentum renormalization scale $\mu=1$ in the present section. We thus get, after plugging in the fixed point values,
\begin{equation}
F_f=F_f^0+F_{fb}^{(1)}=\int dk_x~\frac{3\zeta(3)}{2\pi^2}N_cN_f T^{3-\epsilon} \left(1+(z-1)\frac{T^{-\epsilon}}{\epsilon}\right) \rightarrow \int dk_x~\frac{3\zeta(3)}{2\pi^2}N_cN_f T^{3-\epsilon - (z-1)},
\end{equation}
where the pure $1/\epsilon$ pole is cancelled by the usual addition of a counter term. Similarly,
\begin{equation}
F_b=F_b^0+F_{fb}^{(2)}=\frac{\pi^2}{90c^2}(N_c^2-1)T^{4-\epsilon} \left(1+2\frac{z-1}{\epsilon}T^{-\epsilon}\right) \rightarrow \frac{\pi^2}{90c^2}(N_c^2-1)T^{4-\epsilon-2(z-1)}.
\end{equation}

We now observe that the bosonic term $F_b$ is compatible with the behavior $\sim T^{2-\epsilon + 2/z}$ expected from the hyperscaling preserving
scaling dimension in Eq.~(\ref{Fdim}); the agreement holds to first order in $\epsilon$ after recalling that $z-1$ is $\mathcal{O}(\epsilon)$
from Eq.~(\ref{eq:z}). For the fermionic contribution, as in Section~\ref{subsec:kubo} the behavior depends upon the fate of the $k_x$ integral.
As noted at the beginning of the present section, for the low $T$ behavior we should impose an upper cutoff on the integral of order
$T^{1/z}$; then $F_f \sim T^{3-\epsilon - (z-1) + 1/z}$ which also agrees with $\sim T^{2-\epsilon + 2/z}$ to first order in $\epsilon$.
Thus both the bosonic and fermionic contributions to the free energy obey hyperscaling, and the behavior in Eq.~(\ref{hyper}), up to 
logarithms.

As was the case in Section~\ref{sec:jj}, for very small bare velocity $c$, and for $c\Lambda < T < \Lambda $, there is a regime of hyperscaling violation
when the $k_x$ integral is replaced by $\Lambda$, and behavior is as in Eq.~(\ref{hyperv}). Note that we are using units in which
the velocity $v_F$ in Eq.~(\ref{eq:dispersions}) has been set equal to unity; so the full condition for this intermediate
regime is $c \Lambda < T < v_F \Lambda$.

\section{Quantum Boltzmann Equation}	
\label{sec:boltzmann}

We now compute the hot spot conductivity $\sigma_Q$ appearing in Eq.~(\ref{sQ}) 
in $d=2$ using a quantum Boltzmann equation approach \cite{damle97,SS98,piazza14,kamenev11}. 
We use the Keldysh formalism at one-loop order to derive quantum kinetic equations for the fermions and bosons in the presence of 
an applied electric field, and then solve these equations in linear response to obtain the contribution of the fermions near the 
hot spots to the DC conductivity. Note that, unlike the previous sections, we are not performing a systematic $\epsilon$
expansion here, but working directly in $d=2$ to minimize technical complexity.

As in Section~\ref{sec:free}, we will restrict our analysis to the case of vanishing $v$ and $c$, when the Fermi surfaces are nested,
and manipulations similar to Eq.~(\ref{eq:quasilocaltrick}) can be applied. With finite $v$ and $c$, as argued in Sections~\ref{subsec:kubo}
and \ref{sec:free}, we can estimate the low $T$ hot spot conductivity by limiting the $k_x$ integral along the Fermi surface
by an upper bound of order $T^{1/z}$.

\subsection{Keldysh framework}

\begin{figure}[h]
\includegraphics[width=165mm]{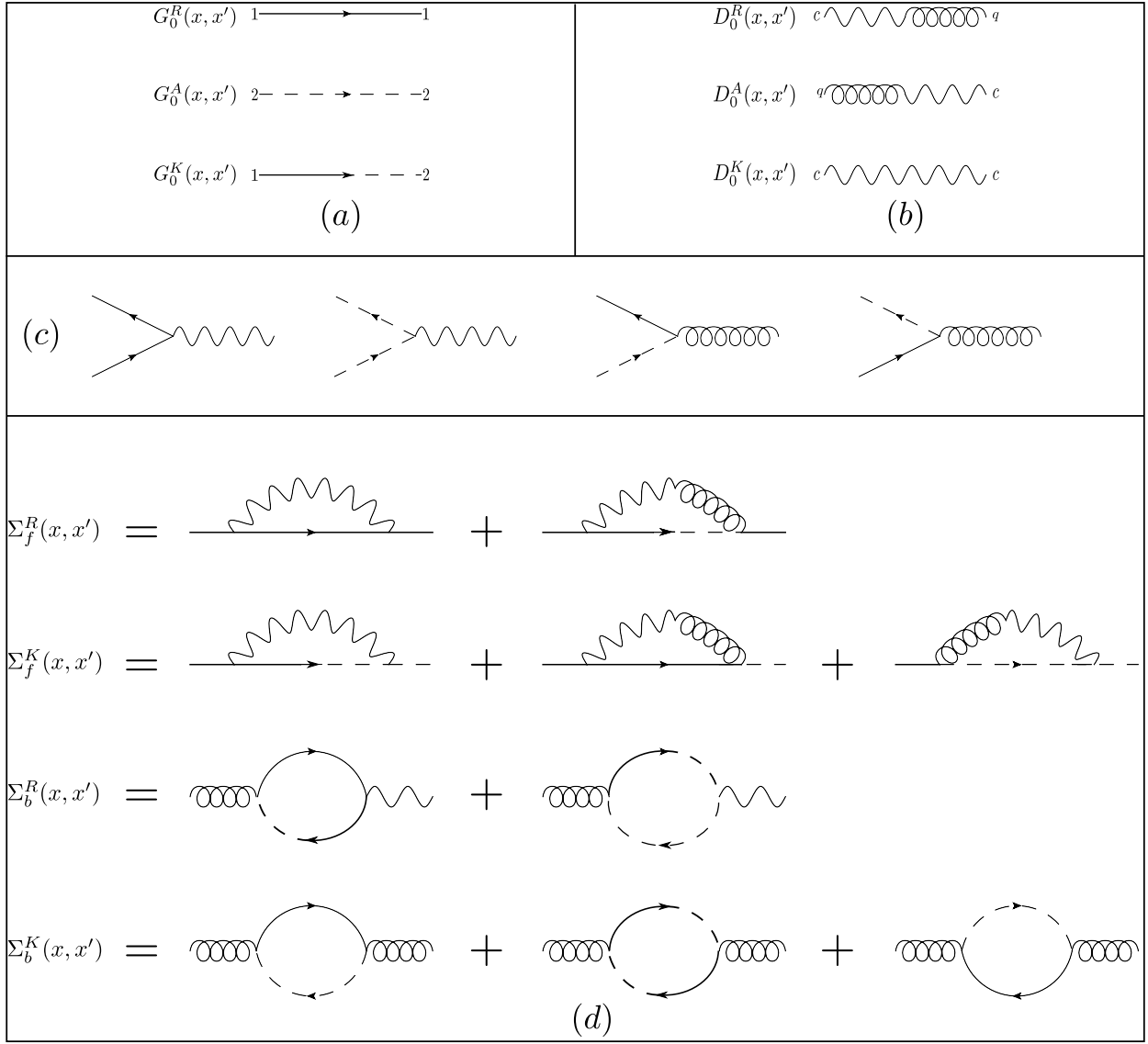}
\caption{Feynman rules and one-loop graphs for the self energies of the fermions and bosons in the Keldysh formalism: (a) Fermion propagators. (b) Boson propagators. (c) Yukawa vertices. (d) Self energies. Here $x=(t,\mathbf{r})$, hot spot indices $(\ell,\pm)$ are suppressed and the external legs on the self energy diagrams are amputated. The external momentum and frequency on the self energy diagrams is on shell. The $2$-$1$ and $q$-$q$ propagators are zero and hence are omitted.}
\label{fig:keldyshgraphs}
\end{figure}

We begin by expressing the action in Eq.~(\ref{eq:action}) on the closed time Keldysh contour \cite{piazza14, kamenev11}. Denoting with subscripts $+$ the forward part of the contour  and with subscripts $-$ the backward part of the contour, we obtain for the free part of the action
\begin{align}
&S_{\bar{\psi}\psi} \nonumber \\
&= \int_{-\infty}^{\infty} d t \int \frac{ d^2 \mathbf{p}} {(2\pi)^2}
\sum_{\ell=1}^{4} \sum_{m=\pm} \sum_{\sigma=\uparrow,\downarrow}
\Bigg[
\bar{\psi}_{\ell,\sigma,+}^{(m)}(t,\mathbf{p}) \left(i \partial_t - e^m_\ell (\mathbf{p})\right) \psi_{\ell,\sigma,+}^{(m)} (t,\mathbf{p}) 
- \bar{\psi}_{\ell,\sigma,-}^{(m)}(t,\mathbf{p}) \left(i \partial_t - e^m_\ell (\mathbf{p})\right) \psi_{\ell,\sigma,-}^{(m)}(t,\mathbf{p})
\Bigg],
\nonumber \\
&S_{\vec{\phi}\vec{\phi}} =\frac{1}{2} \int_{-\infty}^{\infty} d t \int \frac{ d^2 \mathbf{q}} {(2\pi)^2}
\Bigg[
\vec{\phi}_{+}(t,-\mathbf{q})\cdot
\left(
-\partial_t^2
-
\omega_\mathbf{q}^2 
\right)
\vec{\phi}_{+}(t,\mathbf{q})
-
\vec{\phi}_{-}(t,-\mathbf{q})\cdot
\left(
-\partial_t^2
-
\omega_\mathbf{q}^2
\right)
\vec{\phi}_{-}(t,\mathbf{q})
\Bigg],
\label{eq:spinfermion_plusminus}
\end{align}
with $\omega_\mathbf{q}=c|\mathbf{q}|$. The interacting part is given by
\begin{align}
S_{\vec{\phi}\bar{\psi}\psi}
&=
-g \int_{-\infty}^{\infty} d t 
\int d^{2} \mathbf{r}
\sum_{\ell=1}^4 \sum_{\sigma,\sigma'=\uparrow,\downarrow}
\Bigg[
\vec{\phi}_{+}(t,\mathbf{r})
\cdot
\bar{\psi}_{\ell,\sigma,+}^{(+)}(t,\mathbf{r})
\vec{\tau}_{\sigma,\sigma'}
\psi_{\ell,\sigma',+}^{(-)}(t,\mathbf{r})
- \nonumber \\
&-
\vec{\phi}_{-}(t,\mathbf{r})
\cdot
\bar{\psi}_{\ell,\sigma,-}^{(+)}(t,\mathbf{x})\vec{\tau}_{\sigma,\sigma'}(t,\mathbf{r})
\psi_{\ell,\sigma',-}^{(-)}(t,\mathbf{r})
+ h.c.
\Bigg],
\label{eq:yukawa_plusminus}
\end{align}
We now perform the standard bosonic and fermionic Keldysh rotations: For the real bosons we use
\begin{align}
\vec{\phi}_{+}& =  \vec{\phi}_c + \vec{\phi}_q,
\nonumber\\
\vec{\phi}_{-}& =  \vec{\phi}_c - \vec{\phi}_q
\end{align}
and for the Grassmannian fermions we have
\begin{align}
\psi^{(m)}_{\ell,\sigma,+} &= \frac{1}{\sqrt{2}}(\psi_{\ell,\sigma,1}^{(m)} + \psi_{\ell,\sigma,2}^{(m)}),
\quad\quad
\bar{\psi}^{(m)}_{\ell,\sigma,+} = \frac{1}{\sqrt{2}}(\bar{\psi}^{(m)}_{\ell,\sigma,1} + \bar{\psi}^{(m)}_{\ell,\sigma,2}),
\nonumber\\
\psi^{(m)}_{\ell,\sigma,-} &= \frac{1}{\sqrt{2}}(\psi^{(m)}_{\ell,\sigma,1}  - \psi^{(m)}_{\ell,\sigma,2}),
\quad\quad
\bar{\psi}^{(m)}_{\ell,\sigma,-} = \frac{1}{\sqrt{2}}(\bar{\psi}^{(m)}_{\ell,\sigma,2} - \bar{\psi}^{(m)}_{\ell,\sigma,1}).
\end{align}
Hence, we get for the free fermion part of the lagrangian 
\begin{align}
\mathcal{L}_{\bar{\psi}\psi}
=
\sum_{\ell=1}^{4} \sum_{m=\pm} \sum_{\sigma=\uparrow,\downarrow} \left(\bar{\psi}^{(m)}_{\ell,\sigma,1}(t,\mathbf{p})\;  \bar{\psi}^{(m)}_{\ell,\sigma,2}(t,\mathbf{p})\right)
\left( \begin{array}{cc}
\left[G_0^{R\ell m}\right]^{-1} & \delta^{K}_f \\
0                        & \left[G_0^{A\ell m}\right]^{-1}
\end{array} \right)
\left( 
\begin{array}{c}
\psi^{(m)}_{\ell,\sigma,1}(t,\mathbf{p})
\\
\psi^{(m)}_{\ell,\sigma,2}(t,\mathbf{p})
\end{array} 
\right)\;,
\end{align}
where the infinitesimal $\delta_f^{K}$ ensures convergence.  Inverting this matrix, we obtain the bare fermion Green's function matrix
\begin{align}
\widehat{G}_0^{\ell m} 
=
\left( \begin{array}{cc}
G_0^{R\ell m}  & G_0^{K\ell m} \\
0             & G_0^{A\ell m}
\end{array} \right).
\label{eq:fermionfree}
\end{align}
The bare retarded (R) and advanced (A) fermion Green's functions thus are
\begin{align}
G_0^{R\ell m}(\omega,\mathbf{p}) &= \frac{1}{\omega + i 0_{+} - e^m_\ell(\mathbf{p})},
\nonumber\\
G_0^{A\ell m}(\omega,\mathbf{p}) &= \frac{1}{\omega - i 0_{+} - e^m_\ell(\mathbf{p})}.
\end{align}
For the free boson part of the lagrangian we have
\begin{align}
\mathcal{L}_{\vec{\phi}\vec{\phi}}
=
\frac{1}{2}
\left(
\vec{\phi}_c(-\omega,-\mathbf{q})
\;  
\vec{\phi}_q(-\omega,-\mathbf{q})
\right)
\left( \begin{array}{cc}
0                                      & \left[D_0^{A}\right]^{-1}  \\
\left[D_0^{R}\right]^{-1}    &  \delta_b^{K} 
\end{array} \right)
\left( 
\begin{array}{c}
\vec{\phi}_c(\omega,\mathbf{q})
\\
\vec{\phi}_q(\omega,\mathbf{q})
\end{array} 
\right)\;,
\label{eq:bosonfree}
\end{align}
where the infinitesimal $\delta_b^{K}$ again ensures convergence. After performing the matrix inverse,
\begin{align}
\widehat{D}_0 
=
\left( \begin{array}{cc}
D_0^{K}    & D_0^{R} \\
D_0^{A}   & 0
\end{array} \right)
\end{align}
and the retarded and advanced boson Greens' functions hence are
\begin{align}
D_0^{R}(\omega,\mathbf{q}) &= \frac{1}{2}\frac{1}{\left(\omega + i 0_+\right)^2 -\omega_\mathbf{q}^2},
\nonumber\\
D_0^{A}(\omega,\mathbf{q}) &= \frac{1}{2}\frac{1}{\left(\omega - i 0_+\right)^2 -\omega_\mathbf{q}^2}\;.
\end{align}
The interaction between fermions at the $(\ell,+)$ and $(\ell,-)$ hot spots and the boson takes the following form:
\begin{align}
\mathcal{L}_{\vec{\phi}\bar{\psi}\psi}&=
- g\sum_{\ell=1}^4 \sum_{\sigma,\sigma'=\uparrow,\downarrow}
\left(\bar{\psi}^{(+)}_{\ell,\sigma,1}(t,\mathbf{r})\;  \bar{\psi}^{(+)}_{\ell,\sigma,2}(t,\mathbf{r})\right)
\left( \begin{array}{cc}
\vec{\phi}_c(t,\mathbf{r})\cdot\vec{\tau}_{\sigma\sigma'} & \vec{\phi}_q(t,\mathbf{r})\cdot\vec{\tau}_{\sigma\sigma'}  \\
\vec{\phi}_q(t,\mathbf{r})\cdot\vec{\tau}_{\sigma\sigma'}                       &\vec{\phi}_c(t,\mathbf{r})\cdot\vec{\tau}_{\sigma\sigma'} 
\end{array} \right)
\left( 
\begin{array}{c}
\psi^{(-)}_{\ell,\sigma',1}(t,\mathbf{r})
\\
\psi^{(-)}_{\ell,\sigma',2}(t,\mathbf{r})
\end{array} 
\right)+h.c.~\;.
\end{align}
This gives rise to the Feynman rules summarized graphically in Fig.~\ref{fig:keldyshgraphs}.

We adopt the shorthand convention of $x=(t,\mathbf{r})$ and $q=(\omega,\mathbf{q})$ to combine spatial and temporal components. We have the relations
\begin{align}
G^{K\ell m} &= G^{R \ell m} \circ F_f - F_f \circ G^{A \ell m}, \nonumber \\
D^K &= D^R \circ F_b - F_b \circ D^A,
\end{align}
where $C=A\circ B$ implies $C(x,x')=\int dx_1 A(x,x_1)B(x_1,x')$ and $F_{f,b}$ are respectively the fermionic and bosonic distribution functions. The Dyson equations for the matrix fermion and boson Green's functions are
\begin{align}
([\widehat{G}_0^{\ell m}]^{-1}-\widehat{\Sigma}_f^{\ell m})\circ\widehat{G}^{\ell m} &= 1, \nonumber \\
(\widehat{D}_0^{-1}-\widehat{\Sigma}_b)\circ\widehat{D} &= 1.
\end{align}
The self energy matrices $\widehat{\Sigma}$ have the same form as the inverse Green's function matrices in  Eq.~(\ref{eq:fermionfree}) and Eq.~(\ref{eq:bosonfree}), and hence the different components of the self energies are given by the graphs in Fig.~\ref{fig:keldyshgraphs} at the one loop level. Defining central and relative coordinates $x_c=(x+x')/2$ and $x_r=(x-x')/2$, we can convert the two point functions $G$, $D$, $F$ and $\Sigma$ which are of the form $A(x,x')=A(x_c+x_r/2,x_c-x_r/2)$  into a momentum representation via the Wigner transform
\begin{equation}
A(x_c,p)=\int dx_r e^{-ipx_r} A\left(x_c+\frac{x_r}{2},x_c-\frac{x_r}{2}\right).
\end{equation}
Since we have spatial translational invariance in the linear response limit a of weak applied electric field $\mathbf{E}$, we can further simplify $A(x_c,p)\rightarrow A(t,p)$. We will also always consider external particles to be on shell in the subsequent computations of the collision integrals. We define an alternate parameterization $f_{f,b}$ of the distribution functions $F_{f,b}$
\begin{align}
F_f(t,\mathbf{p},\omega) &= 1-2f_f(t,\mathbf{p},\omega), \nonumber \\
F_b(t,\mathbf{q},\omega) &= 1+2f_b(t,\mathbf{q},\omega).
\label{eq:distribother}
\end{align}
In thermal equlibrium in the absence of any applied electric fields, we have $f_{f,b}(t,\mathbf{k},\omega)=n_{f,b}(\omega)$, where $n_{f,b}(\omega)=1/(1\pm e^{\omega/T})$ are the thermal Fermi and Bose functions respectively~\cite{kamenev11}.

\subsection{Kinetic equations for fermions and bosons}

There are two coupled quantum kinetic equations~\cite{kamenev11, fritz11}, one for the electrically charged fermions, 
\begin{equation}
\left(
\frac{\partial}{\partial t} + \mathbf{E} \cdot \frac{\partial}{\partial \mathbf{p}}\right) F_f^{\ell\pm} (t,\mathbf{p})
= I_{f\ell\pm}^{\rm{coll}}[F_f,F_b](t,\mathbf{p}),
\label{eq:fke}
\end{equation}
with the on shell fermion distribution function $F_f^{\ell\pm}(t,\mathbf{p})=F_f(t,\mathbf{p},e_{\ell}^{\pm}(\mathbf{p}))$; and one for the neutral bosons,
\begin{equation}
\frac{\partial}{\partial t} F_b(t,\mathbf{q},\omega_{\mathbf{q}}) = I_b^{\rm{coll}}[F_f,F_b](t,\mathbf{q}),
\label{eq:bke}
\end{equation}
The fermion electric charge is set to 1 for simplicity. The two collision integrals have the general form~\cite{kamenev11}
\begin{align}
I_{f\ell\pm}^{\rm{coll}}[F_f,F_b](t,\mathbf{p}) &= i \Sigma_f^{K\ell\pm}(t,\mathbf{p},e^{\ell\pm}(\mathbf{p})) + 2 F_f^{\ell\pm}(t,\mathbf{p}) {\rm Im}\left[\Sigma_f^{R\ell\pm}(t,\mathbf{p},e^{\ell\pm}(\mathbf{p}))\right],
\nonumber\\
I_b^{\rm{coll}}[F_f,F_b](t,\mathbf{q}) &= i \Sigma_b^K(t,\mathbf{q},\omega_{\mathbf{q}}) + 2 F_b(t,\mathbf{q},\omega_{\mathbf{q}}) {\rm Im}\left[\Sigma_b^R(t,\mathbf{q},\omega_{\mathbf{q}})\right].
\label{eq:coll}
\end{align}
At the one-loop level, $\Sigma^{R,K\ell\pm}_{f,b}$ are given by the graphs in Fig.~\ref{fig:keldyshgraphs}. The self energies and collision integrals are computed in Appendix~\ref{app:Boltzmann}. We obtain
\begin{align}
&I^{\rm coll}_{f\ell\pm}[f_f,f_b](t,\mathbf{p})= \nonumber \\
&=3g^2\int\frac{d^2\mathbf{q}}{2\pi}\frac{1}{\omega_{\mathbf{q}}}\Bigg(\delta(e^{\pm}_{\ell}(\mathbf{p}) - e^{\mp}_{\ell}(\mathbf{p}-\mathbf{q}) - \omega_{\mathbf{q}})\Big[f_f^{\ell\pm}(t,\mathbf{p})(1-f_f^{\ell\mp}(t,\mathbf{p}-\mathbf{q}))-f_f^{\ell\mp}(t,\mathbf{p}-\mathbf{q})f_b(t,\mathbf{q},\omega_{\mathbf{q}})+ \nonumber \\
&+f_f^{\ell\pm}(t,\mathbf{p})f_b(t,\mathbf{q},\omega_{\mathbf{q}})\Big] -\delta(e^{\pm}_{\ell}(\mathbf{p}) -e^{\mp}_{\ell}(\mathbf{p}-\mathbf{q}) + \omega_{\mathbf{q}})\Big[f_f^{\ell\pm}(t,\mathbf{p})(1-f_f^{\ell-}(t,\mathbf{p}-\mathbf{q}))-\nonumber \\
&-f_f^{\ell\mp}(t,\mathbf{p}-\mathbf{q})f_b(t,\mathbf{q},-\omega_{\mathbf{q}})+f_f^{\ell\pm}(t,\mathbf{p})f_b(t,\mathbf{q},-\omega_{\mathbf{q}})\Big]\Bigg), 
\label{eq:fci}
\end{align}
and
\begin{align}
&I^{\rm coll}_{b}[f_f,f_b](t,\mathbf{q})=4g^2\sum_{\ell} \int\frac{d^2\mathbf{k}}{2\pi}\Big[\delta\left(e^-_\ell(\mathbf{k})+ \omega_{\mathbf{q}}-e^+_\ell(\mathbf{k}+\mathbf{q})\right)\Big(f_{f}^{\ell+}(t,\mathbf{k}+\mathbf{q})(1-f_{f}^{\ell-}(t,\mathbf{k}))+ \nonumber \\
&+ f_{f}^{\ell+}(t,\mathbf{k}+\mathbf{q})f_b(t,\mathbf{q},\omega_{\mathbf{q}})-f_{f}^{\ell-}(t,\mathbf{k})f_b(t,\mathbf{q},\omega_{\mathbf{q}})\Big) + (+\leftrightarrow-)\Big], 
\label{eq:bci}
\end{align}
where we have expressed the collision integrals in the alternate parameterization (\ref{eq:distribother}) of the distribution functions $F_{f,b}$ and $f_f^{\ell\pm}(t,\mathbf{p})=f_f(t,\mathbf{p},e_\ell^{\pm}(\mathbf{p}))$.

\subsection{Ansatz and solution for conductivity}

If we set the collision integrals to zero, the distribution function for the neutral bosons unaffected by the applied electric field remains fixed at its equilibrium value. For the Fermions, we have
\begin{align}
\left(
\frac{\partial}{\partial t} + \mathbf{E} \cdot \frac{\partial}{\partial \mathbf{p}}\right) f_{f}^{\ell\pm} (t,\mathbf{p})
&=0.
\label{eq:fkenc}
\end{align}
To solve this in linear response, we switch from the time to the frequency domain and parameterize the deviation of $f_f^{\ell\pm}(\omega,\mathbf{k})$ from its equilibrium value by~\cite{fritz11, fritz08} 
\begin{equation}
f_f^{\ell\pm}(\omega, \mathbf{p})=2\pi\delta(\omega)n_f(\mathbf{v}_\ell^\pm\cdot\mathbf{p})+\mathbf{v}_\ell^\pm\cdot\mathbf{E}(\omega) \varphi(\mathbf{v}^\pm_\ell\cdot\mathbf{p},\omega)n_f(\mathbf{v}_\ell^\pm\cdot\mathbf{p})(1-n_f(\mathbf{v}_\ell^\pm\cdot\mathbf{p})).
\label{eq:fermionpar}
\end{equation}
Inserting this parameterization into Eq.~(\ref{eq:fkenc}), we obtain the collisionless $\varphi$ function in linear response
\begin{equation}
\varphi_{\rm nc}(\mathbf{v}^\pm_\ell\cdot\mathbf{p},\omega)=\frac{1/T}{-i\omega+0^+}.
\end{equation}
We have the electrical current density:
\begin{equation}
\mathbf{J}(\omega)=2\sum_{\ell=1}^4\sum_{m=\pm}\int\frac{d^2\mathbf{p}}{(2\pi)^2}\mathbf{v}_{\ell}^m\langle\psi_\ell^{m\dagger}\psi_\ell^{m}\rangle(\omega,\mathbf{p})=2\sum_{\ell=1}^4\sum_{m=\pm}\int\frac{d^2\mathbf{p}}{(2\pi)^2}\mathbf{v}_{\ell}^m f_f^{\ell m}(\omega,\mathbf{p}),
\end{equation}
and hence obtain the linear response conductivity
\begin{equation}
\sigma_{ij}(\omega)=\frac{\delta J_i(\omega)}{\delta E_j(\omega)}=2\sum_{\ell=1}^4\sum_{m=\pm}\int\frac{d^2\mathbf{p}}{(2\pi)^2}(\mathbf{v}_{\ell}^m\cdot\mathbf{\hat{e}}_i)(\mathbf{v}_{\ell}^m\cdot\mathbf{\hat{e}}_j)\varphi(\mathbf{v}^\pm_\ell\cdot\mathbf{p},\omega)n_f(\mathbf{v}_\ell^\pm\cdot\mathbf{p})(1-n_f(\mathbf{v}_\ell^\pm\cdot\mathbf{p}))).
\label{eq:boltzmannsigma}
\end{equation}

It is easily seen that if the $\varphi_{\rm nc}$ is used in the above expression, we obtain a temperature independent collisionless conductivity that has a delta function in $\omega$ in its real part. In fact, for the Sur-Lee embedding in higher dimensions, we have for the collisionless conductivity
\begin{equation}
\mathrm{Re}[\sigma_{xx}^{\rm nc}(\omega)]=\mathrm{Re}[\sigma_{yy}^{\rm nc}(\omega)]\propto \int dk_x~T^{1-\epsilon}\delta(\omega),
\end{equation}
which is derived in Appendix~\ref{app:Boltzmann}; assuming the $k_x$ integral yields a factor of the cutoff $\Lambda$, this yields a conductivity
with the hyperscaling violating scaling dimension in Eq.~(\ref{Fsdimv}), as expected for free fermions. Once collisions of the fermions with the bosons are included, these delta functions are broadened, and the $k_x$ integral has to be performed with more care, as in Section~\ref{subsec:kubo}.

Returning to $d=2$, we find that, to linear order in $\mathbf{E}$, the bosons still remain in equilibrium and their distribution function is hence given by the thermal Bose function $f_b(t,\mathbf{k},\omega_{\mathbf{k}})=n_b(\omega_{\mathbf{k}})$ if the parameterization Eq.~(\ref{eq:fermionpar}) is used (See Appendix~\ref{app:Boltzmann} for a derivation of this fact). Intuitively, this is because the linearly dispersing hot spot model exhibits particle-hole symmetry, making the charge carrying modes excited by the applied electric field particle-hole pairs with the particle and hole moving in opposite directions. The bosons then do not absorb any momentum that they have to dissipate when the particle-hole pairs recombine and hence remain in equilibrium. This behavior is also present for quantum critical transport in graphene \cite{fritz11}.

We now consider the system to be at the fixed point discussed previously, in the spirit of~\cite{damle97}. We take the applied electric field to be in the $y$ direction ($\mathbf{E}=E_y\mathbf{\hat{e}}_y$); since $v\rightarrow0$ at the fixed point, only the $\ell=1$ and $\ell=3$ pairs of hot spots contribute significant response in this case. (For the electric field in the $x$ direction we obtain the same response with the $\ell=4$ and $\ell=2$ hot spot pairs respectively). We insert the $f_f^{\ell^\pm}$ functions parameterized by Eq.~(\ref{eq:fermionpar}) and the thermal Bose function for $f_b$ into the frequency domain version of the fermion kinetic equation Eq.~(\ref{eq:fke}), and linearize in $E_y$ to get the following integral equation for $\varphi$ in the $v\rightarrow0$ limit (considering the $\ell=1$ pair of hot spots)

\begin{align}
&-\frac{1}{2}\frac{3g^2}{2\pi}\int \frac{d^2\mathbf{q}}{c|\mathbf{q}|}\Bigg(\delta(2p_y-q_y-c|\mathbf{q}|) \Big[\varphi(p_y, \omega)n_f(p_y)\left(1-n_f(p_y)\right)\left(1-n_f(q_y-p_y)+n_b(c|\mathbf{q}|)\right) + \nonumber \\
&+\varphi(q_y-p_y,\omega)n_f(q_y-p_y)\left(1-n_f(q_y-p_y)\right)\left(n_f(p_y)+n_b(c|\mathbf{q}|)\right)\Big] - \nonumber \\
&-\delta(2p_y-q_y+c|\mathbf{q}|)\Big[\varphi(p_y,\omega)n_f(p_y)\left(1-n_f(p_y)\right)\left(1-n_f(q_y-p_y)+n_b(-c|\mathbf{q}|)\right) + \nonumber \\
&+\varphi(q_y-p_y,\omega)n_f(q_y-p_y)\left(1-n_f(q_y-p_y)\right)\left(n_f(p_y)+n_b(-c|\mathbf{q}|)\right)\Big]\Bigg) \nonumber \\
&=(-i\omega+0^+)\varphi(p_y,\omega)n_f(p_y)\left(1-n_f(p_y)\right)-\frac{1}{T}n_f(p_y)\left(1-n_f(p_y)\right). \nonumber \\
\end{align}
In the collision term, the boson momentum parallel to the Fermi surface, $q_x$, is limited by the Bose function to a value of order $T/c$.
Integrating out $q_x$, we obtain
\begin{equation}
\frac{3g^2}{2\pi c}\left[C_0(p_y) \varphi(p_y, \omega)+\frac{C_1[\varphi,p_y]}{n_f(p_y)\left(1-n_f(p_y)\right)}\right]=(-i\omega+0^+)\varphi(p_y,\omega)-\frac{1}{T}.
\label{eq:C0}
\end{equation}
where,
\begin{align}
&C_0(p_y)=\frac{1}{2}\int_{-\infty}^{\infty} dq_y \Bigg[\mathrm{sgn}(q_y- 2p_y)\frac{\Theta \left((q_y-2 p_y)^2-c^2 q_y^2\right)}{\sqrt{(q_y-2 p_y)^2-c^2 q_y^2}} \left(1-n_f(q_y-p_y)+n_b(2p_y-q_y)\right) \Bigg], \nonumber \\
&C_1[\varphi,p_y]= \frac{1}{2}\int_{-\infty}^{\infty} dq_y \Bigg[\mathrm{sgn}(q_y- 2p_y)\frac{\Theta \left((q_y-2 p_y)^2-c^2 q_y^2\right)}{\sqrt{(q_y-2 p_y)^2-c^2 q_y^2}}  \varphi(q_y-p_y,\omega)n_f(q_y-p_y) \times \nonumber \\
&\times \left(1-n_f(q_y-p_y)\right)\left(n_f(p_y)+n_b(2p_y-q_y)\right) \Bigg].
\label{eq:C1}
\end{align}
This equation may be solved iteratively by choosing the trial solution
\begin{align}
\varphi_1(p_y,\omega)=\frac{-1/T}{\frac{3g^2}{2\pi c}C_0(p_y)+i\omega-0^+}
\label{eq:iter1}
\end{align}
and iterating 
\begin{equation}
\left(\frac{3g^2}{2\pi c}C_0(p_y)+i\omega-0^+\right)\varphi_{j+1}(p_y,\omega)=-\frac{1}{T}-\frac{3g^2}{2\pi c}\frac{C_1[\varphi_j,p_y]}{n_f(p_y)\left(1-n_f(p_y)\right)},
\label{eq:iter2}
\end{equation}
for $j>1$ (Note that $\varphi_1$ in Eq.~(\ref{eq:iter1}) may be derived from inserting $\varphi_0=0$ into Eq.~(\ref{eq:iter2})). The integral for $C_1$ may be evaluated numerically by sampling $\varphi$ at a discrete set of points and then constructing an interpolating function through these points. The exponential decay of $n_{f,b}$ at large values of their arguments ensures convergence of the integrals and supresses errors arising from the extrapolation of the interpolating function to large arguments.  The trial solution $\varphi_1$ is actually fairly accurate, and this algorithm converges in a small number of iterations. 

In the limit of $c\rightarrow0$, which also occurs at the fixed point, we see that $C_0\sim1/c$ because the singularity in $n_b(2p_y-q_y)$ as $2p_y-q_y\rightarrow0$ is cut off by $c$ in the $\Theta$ function in Eq.~(\ref{eq:C0}); this also occurs in the integral for $C_1$ in (\ref{eq:C1}). Hence in this limit, we have for $\omega\rightarrow0$,
\begin{equation}
\varphi_1(p_y,0)\approx\frac{c^2}{g^2T}~\chi\left(\frac{p_y}{T}\right)
\end{equation}
for some function $\chi$.
It can be seen from Eq.~(\ref{eq:iter2}), and also established numerically, that given the above form for $\varphi_1$, all $\varphi_{j>1}$ will also be of the same form. Numerically, we find that $\chi$ is an even function and $\chi(0)=0$. Since $\chi$ is an even function, it is easy to show that the $(1,-)$ hot spot contributes the same value to the conductivity as the $(1,+)$ hot spot as it is related by the transformation of $p_y\rightarrow-p_y$ in the above computation. Similarly, the $(3,\pm)$ hot spots also produce identical contributions equal to those from the $(1,\pm)$ hot spots.  Hence, using Eq.~(\ref{eq:boltzmannsigma}),
\begin{equation}
\sigma_{yy}(0)\approx8\frac{c^2}{g^2 T}\int\frac{d^2\mathbf{p}}{(2\pi)^2}\chi\left(\frac{p_y}{T}\right)n_f(p_y)(1-n_f(p_y))) \sim \int dk_x~\frac{c^2}{g^2}. \label{qbe1}
\end{equation}
In the last step we have changed the fermion momentum notation from $p_x$ to $k_x$ for compatibility with earlier discussions.
It is also easily seen that $\sigma_{xx}=\sigma_{yy}$ if we repeat the above analysis for the $\ell=2$ and $\ell=4$ hot spots instead, and that $\sigma_{xy}=\sigma_{yx}=0$. So Eq.~(\ref{qbe1}) is the estimate by the Boltzmann equation of the value of the conductivity
$\sigma_Q$ in Eq.~(\ref{sQ}).

We now need to determine the $T$ dependence implied by Eq.~(\ref{qbe1}) using scaling ideas. Under the renormalization group flow,
we expect that the coupling $\lambda = g^2/v$ flows to a fixed point value. While this fixed point can be determined precisely under an $\epsilon$
expansion, we are only able to make an estimate in the present computation carried out directly in $d=2$, where $g^2 /v$ is a dimensionful
quantity of order $\mu^\epsilon$. The natural scale for the momentum $\mu$ is set by the temperature, and so $\mu \sim T^{1/z}$.
So in $d=2$, we can expect that $g^2 /v \sim T^{1/z}$.
Ignoring the logarithmic factors, we therefore have the estimate
\begin{equation}
\sigma_Q \sim \int dk_x~T^{-1/z}.
\end{equation}
Finally, as in Sections~\ref{sec:jj} and~\ref{sec:free}, we bound the $k_x$ integral by $T^{1/z}$ to conclude that $\sigma_Q \sim \mbox{constant}$, as claimed in Section~\ref{sec:intro}. And also as in previous sections, for a small bare $c$, we will have
$\sigma_Q \sim \Lambda T^{-1/z}$ in the intermediate $T$ regime $c \Lambda < T < \Lambda$ (and as noted at the end of Section~\ref{sec:free}, after restoring units, this condition is $c \Lambda < T < v_F \Lambda$).

\section{Conclusions}
\label{sec:conclusions}

We have computed the critical conductivity and free energy at the onset of spin density wave order in metals in $d=2$
using the $\epsilon$ expansion introduced by Sur and Lee \cite{sur15}.
The advantage of this method is that the $\epsilon$ expansion appears to be valid systematically order-by-order in $\epsilon$,
and there is no breakdown in the renormalization group flows. 
The $\epsilon$ expansion exhibits a logarithmic flow of the velocity $v$ to zero at large length scales, and a dynamic nesting
of the Fermi surfaces. We found that hyperscaling was obeyed, with the hot spot contributions scaling as in Eq.~(\ref{hyper}).

It is interesting to compare these results with a previous
two-loop, large $N$ renormalization group analysis of the spin-density wave critical point in Ref.~\onlinecite{MMSS10b}, which also found a
logarithmic flow of $v$ to zero at low energies. However, it was also found that the large $N$
expansion broke down at sufficiently large scales. The same large $N$ framework was used to compute the optical conductivity by Hartnoll {\em et al.} \cite{HHMS11}, and it was found that hot-spot contribution was $\sigma (\omega)  \sim$ 
constant in the limit of vanishing $v$, as expected under hyperscaling in $d=2$.

This similarity between the large $N$ and $\epsilon$ expansion indicates that the terminology `quasi-local' for the latter expansion \cite{sur15}
should be used with some care, and we have avoided it here. 
The basic scaling properties are similar to those of a standard, spatially-isotropic, critical point obeying hyperscaling
with a finite dynamic critical exponent $z$ given by Eq.~(\ref{eq:z}). The deviations from strong scaling arise only in logarithmic corrections,
which are linked, ultimately, to the asymptotic nesting of the Fermi surfaces \cite{abanov00,MMSS10b} in Fig.~\ref{fig:hotness}b.

We also carried out computations for the free energy density at non-zero $T$ using the $\epsilon$ expansion. Again our results
were in excellent accord with hyperscaling expectations. Both the fermionic excitations at the hot spot and the collective
bosonic fluctuations scaled with the same power of $T$, as shown in Section~\ref{sec:free}.

There was, however, for the somewhat unnatural case of a sufficiently small bare boson velocity, an intermediate energy regime where hyperscaling was violated.
This was discussed in Section~\ref{sec:jj} for the optical conductivity, and in Section~\ref{sec:free} for the free energy. The optical conductivity
results of Refs.~\cite{ACS03,CMY14} are similar to this hyperscaling violating regime, and our analysis indicates that their results do not
apply when the bare boson velocity is not small.

Finally, in Section~\ref{sec:boltzmann}, we addressed the question of the DC conductivity. Because of the constraints of total momentum
conservation, such a computation must be carried out \cite{HKMS,HMPS} in the context of the expression in Eq.~(\ref{sQ}), which separates a quantum critical conductivity $\sigma_Q$ from that associated with `drag' from the conserved momentum. We estimated 
$\sigma_Q$ in Section~\ref{sec:boltzmann} and found a result that scaled as $T^0$,
up to logarithms. Thus, the $\sigma_Q$ contribution to Eq.~(\ref{sQG}), in the theory of the spin density wave critical point, 
is likely not the mechanism of the strange metal linear resistivity.

The momentum-drag term in Eq.~(\ref{sQG}) was considered in a previous work by two of us \cite{AAPSS14} 
for the spin density wave critical point:
there we found that quenched disorder which changes the local critical coupling did lead to a linear-in-$T$ resistivity. This conclusion is 
not modified by the considerations of the present paper.

\section*{Acknowledgments} 

We thank Sung-Sik Lee for helpful explanations of the results of Ref.~\onlinecite{sur15}, and for a discussion
which motivated the analysis in Section~\ref{subsec:kubo}. We are also grateful to
M.~Metlitski and T.~Senthil for valuable discussions. 
This research was supported by the NSF under Grant DMR-1360789, the Templeton foundation, by the Leibniz prize of A. Rosch,
and MURI grant W911NF-14-1-0003 from ARO. Research at Perimeter Institute is supported by the Government of Canada through Industry Canada 
and by the Province of Ontario through the Ministry of Research and Innovation.

\appendix

\section{Computation of $\langle J_y J_y \rangle$}

All computations in this appendix are symbolically described for the $n=1$ fermions, with the identical contributions for $n=3$ accounted for by doubling the overall prefactors. Momentum integrals in dimensional regularization are performed using the standard identity
\begin{equation}
\int d^dk \frac{k^a}{(k^2+\Delta)^b} = \frac{\pi^{d/2}}{\Gamma(d/2)}\frac{\Gamma((a+d)/2)\Gamma(b-(a+d)/2)}{\Gamma(b)}\Delta^{(a+d)/2-b}.
\end{equation}

\subsection{Free fermion contribution to $\langle J_y J_y \rangle$}
\label{app:one-loop}
The free fermion contribution to $\langle J_y J_y \rangle$ is given by Fig.~\ref{fig:JJ_graphs}(a) and is straightforwardly computed in dimensional regularization:
\begin{align}
&\langle J_yJ_y \rangle_{\rm{free}}(\omega)= \nonumber \\
&-2(1+v^2) N_cN_f \int\frac{d^2\mathbf{k}}{(2\pi)^2}\frac{d^{2-\epsilon}\mathbf{K}}{(2\pi)^{2-\epsilon}}\mathrm{Tr}
\left[i\gamma_{d-1}(-i)\frac{\mathbf{\Gamma}\cdot\mathbf{K}+\gamma_{d-1}\varepsilon_1(\mathbf{k})}{\mathbf{K}^2+\varepsilon_1(\mathbf{k})^2}i\gamma_{d-1}(-i)\frac{\mathbf{\Gamma}\cdot(\mathbf{K}+\mathbf{W})+\gamma_{d-1}\varepsilon_1(\mathbf{k})}{(\mathbf{K}+\mathbf{W})^2+\varepsilon_1(\mathbf{k})^2}
\right]  \nonumber \\
&=-4 N_cN_f \sqrt{1+v^2} \int dk_{\parallel} \int_0^1 dx \int\frac{d\varepsilon_1(\mathbf{k})}{(2\pi)^2}\frac{d^{2-\epsilon}\mathbf{K}}{(2\pi)^{2-\epsilon}}\frac{-\mathbf{K}\cdot(\mathbf{K}+\mathbf{W})+\varepsilon_1(\mathbf{k})^2}{[(\mathbf{K}+x\mathbf{W})^2+\varepsilon_1(\mathbf{k})^2+x(1-x)\mathbf{W}^2]^2} \nonumber \\
&=-\frac{N_c N_f}{2\pi}\sqrt{1+v^2}\int dk_\parallel\int_0^1dx \int\frac{d^{2-\epsilon}\mathbf{K}}{(2\pi)^{2-\epsilon}} \left(\frac{-\mathbf{K}^2+\mathbf{W}^2x(1-x)}{[\mathbf{K}^2+x(1-x)\mathbf{W}^2]^{3/2}}+\frac{1}{[\mathbf{K}^2+x(1-x)\mathbf{W}^2]^{1/2}}\right) \nonumber \\
&=-\frac{N_c N_f \pi^{1-\epsilon/2}}{\pi \Gamma(1-\epsilon/2) (2\pi)^{2-\epsilon}}\sqrt{1+v^2}\int dk_\parallel~\omega^{1-\epsilon}\int_0^1 dx\Bigg(\frac{\Gamma(1-\epsilon/2)\Gamma(\epsilon/2-1/2)}{2 \Gamma(1/2)}(x(1-x))^{1/2-\epsilon/2}+ \nonumber \\
&+\frac{\Gamma(1-\epsilon/2)\Gamma(1/2+\epsilon/2)}{2 \Gamma(3/2)} (x(1-x))^{1/2-\epsilon/2}-\frac{\Gamma(2-\epsilon/2)\Gamma(-1/2+\epsilon/2)}{2\Gamma(3/2)}(x(1-x))^{1/2-\epsilon/2}\Bigg) \nonumber \\
&=-\sqrt{1+v^2}\int dk_\parallel~N_cN_f\omega^{1-\epsilon} \left(\frac{1}{16\pi} + \mathcal{O}(\epsilon)\right),
\label{eq:jyjyfree}
\end{align}
where $k_{\parallel}$ is the component of $\mathbf{k}$ along the Fermi surface. 

\subsection{Fermion self-energy correction to $\langle J_y J_y \rangle$}
\label{app:self-one-loop}

In this subsection, we will freely take the limit of vanishing velocities associated with Eq.~(\ref{eq:quasilocaltrick}). The extension to the case of finite velocities will be presented in Appendix~\ref{app:self-one-loop-vcn0}.

This self-energy correction is given by Fig.~\ref{fig:JJ_graphs}(b) and a partner diagram with the boson on the lower fermion line. The sum of the two gives
\begin{align}
&\langle J_y J_y \rangle_{\mathrm{SE}}(\omega) \nonumber \\
&=4 i N_f \int \frac{d^2\mathbf{k}}{(2\pi)^2}\frac{d^{2-\epsilon}\mathbf{K}}{(2\pi)^{2-\epsilon}} \mathrm{Tr}\Bigg[\gamma_{d-1}\frac{\mathbf{\Gamma}\cdot\mathbf{K}+\gamma_{d-1}k_y}{\mathbf{K}^2+k_y^2}\Sigma_1(\mathbf{K},\mathbf{k})\frac{\mathbf{\Gamma}\cdot\mathbf{K}+\gamma_{d-1}k_y}{\mathbf{K}^2+k_y^2} \gamma_{d-1}\frac{\mathbf{\Gamma}\cdot(\mathbf{K}+\mathbf{W})+\gamma_{d-1}k_y}{(\mathbf{K}+\mathbf{W})^2+k_y^2}\Bigg]
\label{eq:fse2l}
\end{align}
We first compute the fermion self energy, given by Fig.~\ref{fig:1-loop}(a):
\begin{align}
&\Sigma_1(\mathbf{K},\mathbf{k}) \nonumber \\
&=\frac{g^2\mu^\epsilon}{N_f}\sum_{j=1}^{N_c^2-1}(\tau^j\tau^j)\int \frac{d^2\mathbf{q}}{(2\pi)^2}\frac{d^{2-\epsilon}\mathbf{Q}}{(2\pi)^{2-\epsilon}}i\gamma_{d-1}(-i)\frac{\mathbf{\Gamma}\cdot(\mathbf{K}+\mathbf{Q})-\gamma_{d-1}(q_y+k_y)}{(\mathbf{K}+\mathbf{Q})^2+(q_y+k_y)^2}i\gamma_{d-1}\frac{1}{\mathbf{Q}^2+c^2 \mathbf{q}^2} \nonumber \\
&=i\frac{g^2\mu^\epsilon}{N_f}\sum_{j=1}^{N_c^2-1}(\tau^j\tau^j)\int_0^1 dx \int \frac{d^2\mathbf{q}}{(2\pi)^2}\frac{d^{2-\epsilon}\mathbf{Q}}{(2\pi)^{2-\epsilon}}\frac{-\mathbf{\Gamma}\cdot\mathbf{K}(1-x)-\gamma_{d-1}(q_y+k_y)}{[\mathbf{Q}^2+ x(1-x)\mathbf{K}^2+c^2\mathbf{q}^2(1-x)+x(q_y+k_y)^2]^2} \nonumber \\
&=i\frac{\pi^{1-\epsilon/2}\Gamma(1+\epsilon/2)}{(2\pi)^{2-\epsilon}}\frac{g^2\mu^\epsilon}{N_f}\sum_{j=1}^{N_c^2-1}(\tau^j\tau^j)\int_0^1 dx \int \frac{d^2\mathbf{q}}{(2\pi)^2}\frac{-\mathbf{\Gamma}\cdot\mathbf{K}(1-x)-\gamma_{d-1}(q_y+k_y)}{[x(1-x)\mathbf{K}^2+c^2\mathbf{q}^2(1-x)+x(q_y+k_y)^2]^{1+\epsilon/2}} \nonumber \\
&=i\frac{\pi^{3/2-\epsilon/2}\Gamma(1/2+\epsilon/2)}{(2\pi)^{4-\epsilon}}\frac{g^2\mu^\epsilon}{cN_f}\sum_{j=1}^{N_c^2-1}(\tau^j\tau^j)\int_0^1 dx \int \frac{dq_y}{\sqrt{1-x}}\frac{-\mathbf{\Gamma}\cdot\mathbf{K}(1-x)-\gamma_{d-1}(q_y+k_y)}{[x(1-x)\mathbf{K}^2+x(q_y+k_y)^2]^{1/2+\epsilon/2}}, \nonumber \\
\end{align}
where in the last step we integrated out $q_x$ and then sent $c\rightarrow0$. After shifting $q_y$ by $k_y$ we integrate it out to get
\begin{equation}
\Sigma_1(\mathbf{K},\mathbf{k})=-i\frac{\pi^{2-\epsilon/2}\Gamma(\epsilon/2)}{(2\pi)^{4-\epsilon}}\frac{g^2\mu^\epsilon}{cN_f}\sum_{j=1}^{N_c^2-1}(\tau^j\tau^j)\int_0^1 dx (x(1-x))^{-\epsilon/2} \sqrt{\frac{1-x}{x}}\frac{\mathbf{\Gamma}\cdot\mathbf{K}}{[\mathbf{K}^2]^{\epsilon/2}}.
\label{eq:fse1l}
\end{equation}
Inserting this into the expression for $\langle J_yJ_y \rangle_{\mathrm{SE}}$, 
\begin{align}
&\langle J_y J_y \rangle_{\mathrm{SE}}(\omega) \nonumber \\
&=16 (1-N_c^2) \frac{\pi^{5/2-\epsilon/2}\Gamma(\epsilon/2)\Gamma(1/2-\epsilon/2)}{2^{1-\epsilon}(2\pi)^{8-2\epsilon}\Gamma(1-\epsilon/2)}\frac{g^2\mu^\epsilon}{c}\int dk_x \int dk_y d^{2-\epsilon}\mathbf{K} \frac{\mathbf{K}^4+\mathbf{W}\cdot\mathbf{K}~\mathbf{K}^2-k_y^2(3\mathbf{K}^2+\mathbf{K}\cdot\mathbf{W})}{(\mathbf{K}^2+k_y^2)^2((\mathbf{K}+\mathbf{W})^2+k_y^2)(\mathbf{K}^2)^{\epsilon/2}}
\nonumber \\
&= 32(1-N_c^2) \frac{\pi^{5/2-\epsilon/2}\Gamma(\epsilon/2)\Gamma(1/2-\epsilon/2)}{2^{1-\epsilon}(2\pi)^{8-2\epsilon}\Gamma(1-\epsilon/2)}\frac{g^2\mu^\epsilon}{c} \int dk_x \int_0^1 dy (1-y) \times \nonumber \\
&\times \int dk_y \frac{d^{2-\epsilon}\mathbf{K}}{(\mathbf{K}^2)^{\epsilon/2}} \frac{\mathbf{K}^4+\mathbf{W}\cdot\mathbf{K}~\mathbf{K}^2-k_y^2(3\mathbf{K}^2+\mathbf{K}\cdot\mathbf{W})}{[(\mathbf{K}+y\mathbf{W})^2+\mathbf{W}^2y(1-y)+k_y^2]^3} \nonumber \\
&= 4(1-N_c^2) \frac{\pi^{7/2-\epsilon/2}\Gamma(\epsilon/2)\Gamma(1/2-\epsilon/2)}{2^{1-\epsilon}(2\pi)^{8-2\epsilon}\Gamma(1-\epsilon/2)}\frac{g^2\mu^\epsilon}{c}\int dk_x \int_0^1 dy (1-y) \times \nonumber \\
&\times\frac{d^{2-\epsilon}\mathbf{K}}{(\mathbf{K}^2)^{\epsilon/2}} \Bigg[3\frac{\mathbf{K}^4+\mathbf{W}\cdot\mathbf{K}~\mathbf{K}^2}{[(\mathbf{K}+y\mathbf{W})^2+\mathbf{W}^2y(1-y)]^{5/2}}- \frac{3\mathbf{K}^2+\mathbf{K}\cdot\mathbf{W}}{[(\mathbf{K}+y\mathbf{W})^2+\mathbf{W}^2y(1-y)]^{3/2}}\Bigg] \nonumber \\
&= 4(1-N_c^2) \frac{\pi^{7/2-\epsilon/2}\Gamma(1/2-\epsilon/2)}{2^{1-\epsilon}(2\pi)^{8-2\epsilon}\Gamma(1-\epsilon/2)}\frac{g^2\mu^\epsilon}{c}\int dk_x \int_0^1 dz \int_0^1 dy (1-y) d^{2-\epsilon}\mathbf{K}\Bigg[ \nonumber \\
&3\frac{\mathbf{K}^4+\mathbf{W}\cdot\mathbf{K}~\mathbf{K}^2}{[(\mathbf{K}+yz\mathbf{W})^2+y^2z(1-z)\mathbf{W}^2+\mathbf{W}^2yz(1-y)]^{5/2+\epsilon/2}}\frac{\Gamma(5/2+\epsilon/2)}{\Gamma(5/2)}z^{3/2}(1-z)^{\epsilon/2-1}- \nonumber \\
&- \frac{3\mathbf{K}^2+\mathbf{K}\cdot\mathbf{W}}{[(\mathbf{K}+yz\mathbf{W})^2+y^2z(1-z)\mathbf{W}^2+\mathbf{W}^2yz(1-y)]^{3/2+\epsilon/2}}\frac{\Gamma(3/2+\epsilon/2)}{\Gamma(3/2)}z^{1/2}(1-z)^{\epsilon/2-1}\Bigg]. 
\end{align}
Now we shift $\mathbf{K}\rightarrow\mathbf{K}-yz\mathbf{W}$. This leads to the replacements ($(\mathbf{K}\cdot\mathbf{W})^2\equiv\mathbf{K}^2\mathbf{W}^2/(2-\epsilon)$ as far as integration over $\mathbf{K}$ is concerned)
\begin{align}
&\mathbf{K}^4+\mathbf{W}\cdot\mathbf{K}~\mathbf{K}^2\rightarrow \mathbf{K}^4+\frac{4 \mathbf{K}^2 \mathbf{W}^2 y^2z^2}{2-\epsilon }+2 \mathbf{K}^2 \mathbf{W}^2 y^2z^2-\frac{2 \mathbf{K}^2 \mathbf{W}^2 yz}{2-\epsilon }- \nonumber \\
&-\mathbf{K}^2 \mathbf{W}^2 yz+\mathbf{W}^4 y^4z^4-\mathbf{W}^4 y^3z^3, \nonumber \\
&\mathbf{K}^2 \rightarrow \mathbf{K}^2 + \mathbf{W}^2 y^2z^2, \nonumber \\
&\mathbf{K}\cdot\mathbf{W}\rightarrow -yz\mathbf{W}^2.
\end{align}
Then, integrating out $\mathbf{K}$,
\begin{align}
&\langle J_y J_y \rangle_{\mathrm{SE}}(\omega) = 4(1-N_c^2)\frac{\pi^{9/2-\epsilon}\Gamma(1/2-\epsilon/2)}{2^{-\epsilon}(2\pi)^{8-2\epsilon}\Gamma(1-\epsilon/2)^2}\frac{g^2\mu^\epsilon}{c} \omega^{1-2\epsilon}\int dk_x \int_0^1 dz \int_0^1 dy \Bigg[(1-y) (1-z)^{\epsilon/2-1} \times \nonumber \\
&\times\Bigg(-\frac{\left(3 y^2 z^2-y z\right) \left(\Gamma \left(\epsilon +\frac{1}{2}\right)\Gamma \left(1-\frac{\epsilon}{2}\right)\right) \left(y^2 (1-z) z+(1-y) y z\right)^{-\epsilon -1/2}z^{1/2}}{2 \Gamma \left(\frac{3}{2}\right)}+ \nonumber \\
&+\frac{3 y z \left(\Gamma \left(\epsilon +\frac{1}{2}\right) \Gamma \left(2-\frac{\epsilon }{2}\right)\right) \left(y z \left(\frac{4}{2-\epsilon }+2\right)-\left(\frac{2}{2-\epsilon }+1\right)\right) \left(y^2 (1-z) z+(1-y) y z\right)^{-\epsilon -1/2}z^{3/2}}{2 \Gamma \left(\frac{5}{2}\right)}- \nonumber \\
&-\frac{3 \left(\Gamma \left(\epsilon -\frac{1}{2}\right) \Gamma \left(2-\frac{\epsilon }{2}\right)\right) \left(y^2 (1-z) z+(1-y) y z\right)^{1/2-\epsilon}z^{1/2}}{2 \Gamma \left(\frac{3}{2}\right)}+ \nonumber \\
&+\frac{3 \left(\Gamma \left(\epsilon -\frac{1}{2}\right) \Gamma \left(3-\frac{\epsilon }{2}\right)\right) \left(y^2 (1-z) z+(1-y) y z\right)^{1/2-\epsilon}z^{3/2}}{2 \Gamma \left(\frac{5}{2}\right)}+\nonumber \\
&+\frac{3 y^3 z^3 (y z-1) \left(\Gamma \left(\epsilon +\frac{3}{2}\right) \Gamma \left(1-\frac{\epsilon }{2}\right)\right) \left(y^2 (1-z) z+(1-y) y z\right)^{-\epsilon -3/2}z^{3/2}}{2 \Gamma \left(\frac{5}{2}\right)}\Bigg)\Bigg]. 
\end{align}
To leading order in $\epsilon$, we can take only the $(1-z)^{\epsilon/2-1}$ term in the above integrand for its $z$ dependence and set $z=1$ elsewhere (which produces $2/\epsilon$ for the integral over $z$). This agrees with numerically evaluating the $y$ and $z$ integrals. We thus get,
\begin{align}
&\langle J_y J_y \rangle_{\mathrm{SE}}(\omega)=\int dk_x \frac{(1-N_c^2)}{32\pi^3\epsilon}\frac{g^2\mu^\epsilon}{c} \omega^{1-2\epsilon} \int_0^1 dy~y(6y-5) \sqrt{\frac{1-y}{y}} \nonumber \\
&=\int dk_x (N_c^2-1) \frac{g^2\mu^\epsilon}{c}\omega^{1-2\epsilon}\left(\frac{1}{128 \pi ^2 \epsilon}+\mathcal{O}(1)\right).
\label{eq:JJSEpole}
\end{align}

\subsection{Vertex correction to $\langle J_y J_y \rangle$}
\label{app:vert-one-loop}

As in Appendix~\ref{app:self-one-loop}, here too we will freely take the limit of vanishing velocities associated with Eq.~(\ref{eq:quasilocaltrick}). The case of finite velocities will be presented in Appendix~\ref{app:self-one-loop-vcn0}.

This correction is then given by Fig.~\ref{fig:JJ_graphs}(c):
\begin{align}
&\langle J_y J_y \rangle_{\rm vert}(\omega) = 2 i N_f \int\frac{d^2\mathbf{k}}{(2\pi)^2}\int\frac{d^{2-\epsilon}\mathbf{K}}{(2\pi)^{2-\epsilon}}\mathrm{Tr}\Bigg[\gamma_{d-1}\frac{\mathbf{\Gamma}\cdot\mathbf{K}+\gamma_{d-1}k_y}{\mathbf{K}^2+k_y^2}\Xi_3(\mathbf{K},\mathbf{k},\mathbf{W})\frac{\mathbf{\Gamma}\cdot(\mathbf{K}+\mathbf{W})+\gamma_{d-1}k_y}{(\mathbf{K}+\mathbf{W})^2+k_y^2}\Bigg].
\label{eq:fvx2l}
\end{align}
We again first compute the current ($J_y$) vertex, given by Fig.~\ref{fig:1-loop}(b):
\begin{align}
&\Xi_3(\mathbf{K},\mathbf{k},\mathbf{W}) \nonumber \\
&= \frac{g^2\mu^\epsilon}{N_f}\sum_{j=1}^{N_c^2-1}(\tau^j\tau^j)\int \frac{d^2\mathbf{q}}{(2\pi)^2}\frac{d^{2-\epsilon}\mathbf{Q}}{(2\pi)^{2-\epsilon}}i\gamma_{d-1}\times \nonumber \\
&\times(-i)\frac{\mathbf{\Gamma}\cdot(\mathbf{K}+\mathbf{Q})-\gamma_{d-1}(q_y+k_y)}{(\mathbf{K}+\mathbf{Q})^2+(q_y+k_y)^2}(-i\gamma_{d-1})(-i)\frac{\mathbf{\Gamma}\cdot(\mathbf{K}+\mathbf{Q}+\mathbf{W})-\gamma_{d-1}(q_y+k_y)}{(\mathbf{K}+\mathbf{Q}+\mathbf{W})^2+(q_y+k_y)^2}i\gamma_{d-1}\frac{1}{\mathbf{Q}^2+c^2 \mathbf{q}^2} \nonumber \\ 
&= -2i\frac{g^2\mu^\epsilon}{N_f}\sum_{j=1}^{N_c^2-1}(\tau^j\tau^j)\int_0^1 dx\int_0^{1-x} dy \int \frac{d^2\mathbf{q}}{(2\pi)^2}\frac{d^{2-\epsilon}\mathbf{Q}}{(2\pi)^{2-\epsilon}} \Big[-(\mathbf{K}+\mathbf{Q})\cdot(\mathbf{K}+\mathbf{Q}+\mathbf{W})+(q_y+k_y)^2- \nonumber \\
&+2\mathbf{\Gamma}\cdot(\mathbf{K}+\mathbf{Q})\gamma_{d-1}(q_y+k_y)+\mathbf{\Gamma}\cdot\mathbf{W}\gamma_{d-1}(q_y+k_y)-\mathbf{\Gamma}\cdot(\bar{\mathbf{K}}+\bar{\mathbf{Q}})\mathbf{\Gamma}\cdot\mathbf{W}\Big] \times \nonumber \\
&\times \Big[(\mathbf{Q}+(x+y)\mathbf{K}+y\mathbf{W})^2+\mathbf{W}^2y(1-y)+(1-(x+y))(\mathbf{K}^2(x+y)+2\mathbf{K}\cdot\mathbf{W}y)+ \nonumber \\
&+(x+y)(q_y+k_y)^2+(1-(x+y))\mathbf{q}^2c^2\Big]^{-3}\gamma_{d-1} \nonumber \\
&= -i\frac{g^2\mu^\epsilon\pi^{3/2-\epsilon/2}}{c N_f(2\pi)^{4-\epsilon}\Gamma(1-\epsilon/2)}\sum_{j=1}^{N_c^2-1}(\tau^j\tau^j) \int_0^1 dx \int_0^{1-x} dy \frac{1}{\sqrt{(1-(x+y))}}\int dq_y\Bigg( \nonumber \\
&\Big[(k_y+q_y)^2-(\mathbf{K}(x+y-1)+\mathbf{W}(y-1))\cdot (\mathbf{K}(x+y-1)+\mathbf{W}y)+ \nonumber \\
&+2 \mathbf{\Gamma} \cdot (\mathbf{K} (1-(x+y))-\mathbf{W}y) \gamma_{d-1} (k_y+q_y)-\mathbf{\bar{\Gamma}}\cdot\bar{\mathbf{K}}(1-(x+y))\mathbf{\Gamma}\cdot\mathbf{W}+\mathbf{\Gamma}\cdot\mathbf{W}\gamma_{d-1}(k_y+q_y)\Big]\times \nonumber \\
&\times\Gamma(3/2+\epsilon/2)\Gamma(1-\epsilon/2)\Delta_1^{-(3/2+\epsilon/2)}-\Gamma(1/2+\epsilon/2)\Gamma(2-\epsilon/2)\Delta_1^{-(1/2+\epsilon/2)}\Bigg)\gamma_{d-1},
\end{align}
where we again integrated out $q_x$ and then sent $c\rightarrow0$ in the last step of the above, and
\begin{equation}
\Delta_1=\mathbf{W}^2y(1-y)+(1-(x+y))(\mathbf{K}^2(x+y)+2\mathbf{W}\cdot\mathbf{K}y)+(x+y)(k_y+q_y)^2 \nonumber \\
\end{equation}
Proceeding,
\begin{align}
&\Xi_3(\mathbf{K},\mathbf{k},\mathbf{W})= -i\frac{g^2\mu^\epsilon\pi^{2-\epsilon/2}}{c N_f(2\pi)^{4-\epsilon}\Gamma(1-\epsilon/2)}\sum_{j=1}^{N_c^2-1}(\tau^j\tau^j) \int_0^1 dx \int_0^{1-x} dy \frac{1}{\sqrt{(1-(x+y))(x+y)}}\Bigg( \nonumber \\
&\Bigg[-(\mathbf{K}(x+y-1)+\mathbf{W}(y-1))\cdot (\mathbf{K}(x+y-1)+\mathbf{W}y)- \mathbf{\bar{\Gamma}}\cdot\bar{\mathbf{K}}(1-(x+y))\mathbf{\Gamma}\cdot\mathbf{W} \Bigg]\times \nonumber \\
&\times\frac{\Gamma(1-\epsilon/2)\Gamma(1+\epsilon/2)}{\Delta_2^{(1+\epsilon/2)}}- \frac{\Gamma(\epsilon/2)}{\Delta_2^{\epsilon/2}}\left(\Gamma(2-\epsilon/2)-\frac{\Gamma(1-\epsilon/2)}{2(x+y)}\right)\gamma_{d-1},
\end{align}
where
\begin{equation}
\Delta_2=\mathbf{W}^2y(1-y)+(1-(x+y))(\mathbf{K}^2(x+y)+2\mathbf{W}\cdot\mathbf{K}y).
\end{equation}
An important feature of the above computation is that because
\begin{equation}
\int_0^1 dx\int_0^{1-x} dy \frac{2(x+y)-1}{(x+y)^{3/2}\sqrt{1-(x+y)}}=0,
\end{equation} 
the coefficient of the $1/\epsilon$ pole (i.e. the coefficient of $\Gamma(\epsilon/2)$) in $\Xi$ vanishes when $\epsilon\rightarrow0$. This eventually leads to the lack of a $1/\epsilon$ pole in $\langle J_y J_y \rangle_{\rm vert}$, and hence the correction to scaling of $\langle J_y J_y\rangle$ arises solely from the self-energy graphs.

Taking the expression for the current vertex and inserting it into the one for $\langle J_y J_y \rangle_{\rm vert}$, we get
\begin{align}
&\langle J_y J_y \rangle_{\rm vert}(\omega) \nonumber \\
&=\frac{4 g^2\mu^\epsilon\pi^{3-\epsilon/2}(N_c^2-1)}{c (2\pi)^{8-2\epsilon}\Gamma(1-\epsilon/2)}\int dk_x \int d^{2-\epsilon}\mathbf{K}\int_0^1 dz \int_0^1 dx \int_0^{1-x} \frac{dy}{\sqrt{(1-(x+y))((x+y)}}\Bigg( \nonumber \\
&\Bigg[-(\mathbf{K}(x+y-1)+\mathbf{W}(y-1))\cdot (\mathbf{K}(x+y-1)+\mathbf{W}y)\Bigg(\frac{1}{\Delta_3^{1/2}}- \frac{\mathbf{K}\cdot(\mathbf{K}+\mathbf{W})}{\Delta_3^{3/2}}\Bigg)+ \nonumber \\
&+\frac{\mathbf{\bar{K}}^2\mathbf{W}^2(1-(x+y))}{\Delta_3^{3/2}}\Bigg] \frac{\Gamma(1-\epsilon/2)\Gamma(1+\epsilon/2)}{\Delta_2^{(1+\epsilon/2)}}-\frac{\Gamma(\epsilon/2)}{\Delta_2^{\epsilon/2}}\Bigg(\Gamma(2-\epsilon/2)-\nonumber \\
&-\frac{\Gamma(1-\epsilon/2)}{2(x+y)}\Bigg) \Bigg(\frac{1}{\Delta_3^{1/2}}- \frac{\mathbf{K}\cdot(\mathbf{K}+\mathbf{W})}{\Delta_3^{3/2}}\Bigg)\Bigg),
\end{align}
where now
\begin{align}
&\Delta_2=\mathbf{W}^2y(1-y)+(1-(x+y))(\mathbf{K}^2(x+y)+2\mathbf{W}\cdot\mathbf{K}y), \nonumber \\
&\Delta_3=(\mathbf{K}+z\mathbf{W})^2+z(1-z)\mathbf{W}^2.
\end{align}
We combine denominators using
\begin{equation}
\frac{1}{\Delta_2^s\Delta_3^b}=\frac{\Gamma(s+b)}{\Gamma(s)\Gamma(b)}\int_0^1da \frac{a^{s-1}(1-a)^{b-1}}{[a\Delta_2+(1-a)\Delta_3]^{s+b}},
\end{equation}
and the denominator square completion is
\begin{align}
&a\Delta_2+(1-a)\Delta_3=(a (x+y) (1-(x+y))+(1-a)) \left(\frac{\mathbf{W} (a y (1-(x+y))+(1-a) z)}{a (x+y) (1-(x+y))+(1-a)}+\mathbf{K}\right)^2+ \nonumber \\
&+\mathbf{W}^2 \left(-\frac{(a y (1-(x+y))+(1-a) z)^2}{a (x+y) (1-(x+y))+(1-a)}+a (1-y) y+(1-a) z\right).
\end{align}
Defining
\begin{align}
&f_1=a (x+y) (1-(x+y))+(1-a), \nonumber \\
&f=\frac{1}{f_1}(a y (1-(x+y))+(1-a) z),\nonumber \\
&f_2=a (1-y) y+(1-a) z-f_1 f^2, 
\end{align}
we can process the numerators and write down the final expression 
\begin{equation}
\langle J_y J_y \rangle_{\rm vert}(\omega)=\frac{8 g^2\mu^\epsilon\pi^{4-\epsilon}(N_c^2-1)}{c (2\pi)^{8-2\epsilon}\Gamma(1-\epsilon/2)^2}\omega^{1-2\epsilon}\int dk_x \int_0^1 da \int_0^1 dz \int_0^1 dx \int_0^{1-x} dy\Bigg(T_1+T_2+T_3+T_4\Bigg),
\end{equation}
where
\begin{align}
&T_1=\frac{\sqrt{1-a} a^{\epsilon /2} f_1^{\frac{\epsilon }{2}-3} f_2^{-\epsilon -\frac{3}{2}}}{{\sqrt{-(x+y-1) (x+y)\pi}}}\Gamma \left(1-\frac{\epsilon }{2}\right)\Bigg( \frac{1}{\epsilon-2}f_1 f_2 \Gamma \left(2-\frac{\epsilon }{2}\right) \Gamma \left(\epsilon +\frac{1}{2}\right) \Bigg(2 f^2 (\epsilon -4)(x+y-1)^2- \nonumber \\
&-f (\epsilon -4) (x+3 y-2) (x+y-1)-2 x y+x+(y-1) (y (\epsilon -4)+1)-(1-(x+y))(\epsilon-1)\Bigg)+ \nonumber \\
&+ (f-1) f f_1^2 \Gamma \left(1-\frac{\epsilon }{2}\right) \Gamma \left(\epsilon +\frac{3}{2}\right) (f (x+y-1)-y) (f (x+y-1)-y+1)+ \nonumber \\
&+f_2^2 (x+y-1)^2 \Gamma \left(3-\frac{\epsilon }{2}\right) \Gamma \left(\epsilon -\frac{1}{2}\right)\Bigg), \nonumber \\
&T_2=\frac{a^{\epsilon /2} f_1^{\frac{\epsilon }{2}-2} f_2^{-\epsilon -\frac{1}{2}}}{\sqrt{1-a} \sqrt{-(x+y-1)) (x+y)\pi}} \Gamma \left(1-\frac{\epsilon }{2}\right) \Bigg(-f_1 \Gamma \left(1-\frac{\epsilon }{2}\right) \Gamma \left(\epsilon +\frac{1}{2}\right) (f (x+y-1)-y) \times \nonumber \\
&\times (f (x+y-1)-y+1)-f_2 (x+y-1)^2 \Gamma \left(2-\frac{\epsilon }{2}\right) \Gamma \left(\epsilon -\frac{1}{2}\right)\Bigg), \nonumber \\
&T_3=\frac{\sqrt{1-a} a^{\frac{\epsilon}{2}-1} f_1^{\frac{\epsilon }{2}-2} f_2^{-\epsilon -\frac{1}{2}}}{4 \sqrt{\pi } \sqrt{-x-y+1} (x+y)^{3/2}} \Gamma \left(1-\frac{\epsilon }{2}\right)^2 \Gamma \left(\epsilon -\frac{1}{2}\right) (x (\epsilon -2)+y (\epsilon -2)+1) \Bigg(f^2 \times \nonumber \\
&\times (f_1-2 f_1 \epsilon )+f f_1 (2 \epsilon -1)+f_2 (\epsilon -2)\Bigg), \nonumber \\
&T_4= \frac{a^{\frac{\epsilon }{2}-1} f_1^{\frac{\epsilon }{2}-1} f_2^{\frac{1}{2}-\epsilon }}{4 \sqrt{1-a} \sqrt{\pi  (-x-y+1)} (x+y)^{3/2}} \Gamma \left(1-\frac{\epsilon }{2}\right)^2 \Gamma \left(\epsilon -\frac{1}{2}\right) (x (\epsilon -2)+y (\epsilon -2)+1).
\end{align}
This multidimensional integral over four parameters is finite in the limit of $\epsilon\rightarrow0$ and can be done numerically. We first integrate over $x$ and $y$: The resulting function of $a$ and $z$ has integrable singularities in the limits of $a\rightarrow1$ and $a\rightarrow0$ which can be handled by numerical integration using an adaptive grid. The final result is 
\begin{equation}
\langle J_y J_y \rangle_{\rm vert}(\omega)=\int dk_x \frac{g^2\mu^\epsilon(N_c^2-1)}{32\pi^4 c}\omega^{1-2\epsilon}\Bigg(\alpha_0+\mathcal{O}(\epsilon)\Bigg),
\end{equation}
where $\alpha_0 \approx 1.1$ is a finite numerical constant. 

\section{Free Energy Computations}
\label{app:free-energy}
As in the previous appendix, we will freely take the limit of vanishing velocities associated with Eq.~(\ref{eq:quasilocaltrick}) here as well to compute the correction to the fermion free energy. 
\begin{equation}
F_{fb}=\frac{1}{2}\mathrm{Tr}\left[\sum_{j=1}^{N_c^2-1}\tau^j\tau^j\right]\int\frac{d^2\mathbf{q}d^{1-\epsilon}\mathbf{\bar{Q}}}{(2\pi)^{3-\epsilon}}T\sum_{\omega_q}\frac{1}{\mathbf{\bar{Q}}^2+c^2|\mathbf{q}|^2+\omega_q^2}\left[\left(\Pi(q,T)-\Pi(q,0)\right)+\Pi(q,0)\right].
\label{eq:ffb0}
\end{equation}
Where $\Pi(q,T)$ is the fermion RPA bubble at external momentum and frequency given by $q$ evaluated at temperature $T$. As described in the main text, we evaluate the finite temperature part of the bubble at $v=0$ to renormalize the fermion free energy and the zero temperature part at $v\neq0$ to renormalize the boson free energy. To evaluate the frequency summations, we use the following zeta-function regularization identities:
\begin{align}
T\sum_{\omega_q}\frac{1}{|\omega_q|^s}=2\frac{T^{1-s}}{(2\pi)^s}\zeta(s), \nonumber \\
T\sum_{\omega_k}\frac{1}{|\omega_k|^s}=2\frac{T^{1-s}}{(2\pi)^s}\zeta\left(s,\frac{1}{2}\right).
\end{align}
Where $\omega_q$ is a bosonic Matsubara frequency and $\omega_k$ is a fermionic Matsubara frequency. We then have
\begin{align}
&\Pi(q,T)-\Pi(q,0)=-4g^2\int\frac{dk_x}{2\pi}\int\frac{dk_yd^{1-\epsilon}\mathbf{\bar{K}}}{(2\pi)^{2-\epsilon}}\left(T\sum_{\omega_k}-\int\frac{d\omega_k}{2\pi}\right)\mathrm{Tr}\left[i\gamma_{d-1}G(k)i\gamma_{d-1}G(k+q)\right] \nonumber \\
&=-4 g^2\int dk_x \int\frac{dk_yd^{1-\epsilon}\mathbf{\bar{K}}}{(2\pi)^{3-\epsilon}}\left(T\sum_{\omega_k}-\int\frac{d\omega_k}{2\pi}\right)\Bigg[\frac{\mathbf{\bar{Q}}^2+q_y^2+\omega_q^2}{(\mathbf{\bar{K}}^2+k_y^2+\omega_k^2)((\mathbf{\bar{K}}+\mathbf{\bar{Q}})^2+(k_y+q_y)^2+(\omega_k+\omega_q)^2)}- \nonumber \\
&- \frac{1}{\mathbf{\bar{K}}^2+k_y^2+\omega_k^2}-\frac{1}{(\mathbf{\bar{K}}+\mathbf{\bar{Q}})^2+(k_y+q_y)^2+(\omega_k+\omega_q)^2}\Bigg].
\end{align}
The last two terms in the square brackets yield identical contributions, because the $q$ in the last term can be shifted out. Thus,
\begin{align}
&\Pi(q,T)-\Pi(q,0) \nonumber \\
&=-4g^2\int dk_x \int\frac{dk_yd^{1-\epsilon}\mathbf{\bar{K}}}{(2\pi)^{3-\epsilon}}\left(T\sum_{\omega_k}-\int\frac{d\omega_k}{2\pi}\right)\left[\frac{\mathbf{\bar{Q}}^2+q_y^2+\omega_q^2}{(\mathbf{\bar{K}}^2+k_y^2+\omega_k^2)((\mathbf{\bar{K}}+\mathbf{\bar{Q}})^2+(k_y+q_y)^2+(\omega_k+\omega_q)^2)}\right]+  \nonumber \\
&+8 g^2\int dk_x\int\frac{dk_yd^{1-\epsilon}\mathbf{\bar{K}}}{(2\pi)^{3-\epsilon}}\left(T\sum_{\omega_k}-\int\frac{d\omega_k}{2\pi}\right) \frac{1}{\mathbf{\bar{K}}^2+k_y^2+\omega_k^2}. \nonumber \\
\end{align}
We evaluate the second term to leading order in $\epsilon$ in the above using dimensional regularization for the momentum integral and zeta function regularization for the frequency sum:
\begin{align}
&8 g^2\int dk_x\int\frac{dk_yd^{1-\epsilon}\mathbf{\bar{K}}}{(2\pi)^{3-\epsilon}}\left(T\sum_{\omega_k}-\int\frac{d\omega_k}{2\pi}\right) \frac{1}{\mathbf{\bar{K}}^2+k_y^2+\omega_k^2} = 8 g^2\int dk_x\int\frac{dk_yd^{1-\epsilon}\mathbf{\bar{K}}}{(2\pi)^{3-\epsilon}}T\sum_{\omega_k}\frac{1}{\mathbf{\bar{K}}^2+k_y^2+\omega_k^2} \nonumber \\
&=\frac{8\pi g^2}{(2\pi)^2\epsilon}\int dk_x T\sum_{\omega_k}\frac{1}{|\omega_k|^{\epsilon}} =-\int dk_x \frac{8 g^2 T^{1-\epsilon} \ln 2}{(2\pi)^2},
\end{align}
where we have used the fact that scaleless integrals vanish in dimensional regularization. Thus,
\begin{align}
&\Pi(q,T)-\Pi(q,0) \nonumber \\
&=4 g^2\int dk_x\int\frac{dk_yd^{1-\epsilon}\mathbf{\bar{K}}}{(2\pi)^{3-\epsilon}}\left(\int\frac{d\omega_k}{2\pi}-T\sum_{\omega_k}\right)\left[\frac{\mathbf{\bar{Q}}^2+q_y^2+\omega_q^2}{(\mathbf{\bar{K}}^2+k_y^2+\omega_k^2)((\mathbf{\bar{K}}+\mathbf{\bar{Q}})^2+(k_y+q_y)^2+(\omega_k+\omega_q)^2)}\right]- \nonumber \\
&- \int dk_x\frac{8 g^2 T^{1-\epsilon} \ln 2}{(2\pi)^2},
\end{align}
To evaluate the first term, we introduce a Feynman parameter $y$ to combine the denominators. Doing the $k$ momentum integral and $\omega_k$ frequency summation (integral for the $T=0$ part), we have, to leading order in $\epsilon$ 
\begin{align}
&\Pi(q, T)-\Pi(q, 0) =-\frac{g^2}{8\pi^2}\int dk_x \int_0^1 dy\left[t(y,\tilde{q}^2,\omega_q,\epsilon)-2\pi\right](\tilde{q}^2+\omega_q^2)^{1/2-\epsilon/2} - \int dk_x\frac{8 g^2 T^{1-\epsilon} \ln 2}{(2\pi)^2},
\end{align}
Where $\tilde{q}=(\mathbf{\bar{Q}}^2+q_y^2)^{1/2}$. We determine the following asymptotic expansion numerically
\begin{align}
&\int_0^1 dy~t(y,\tilde{q}^2,\omega_q,\epsilon) =\left(2\pi-T^{1-\epsilon}\frac{16 \ln 2}{(\tilde{q}^2+\omega_q^2)^{1/2-\epsilon/2}}+48\zeta(3)T^{3-\epsilon}\frac{2\omega_q^2-\tilde{q}^2}{(\tilde{q}^2+\omega_q^2)^{5/2-\epsilon/2}}+ O\left(\frac{T^{5-\epsilon}}{\tilde{q}^{5-\epsilon},\omega_q^{5-\epsilon}}\right) + ...\right).
\label{eq:pi}
\end{align}
Simple power counting dictates that the higher terms in the above asymptotic expansion can't produce any log UV divergences in the final two loop graph because they fall off too fast in $q$. Thus, retaining only terms that will survive and contribute to the pole in the final two-loop integral, 
\begin{equation}
\Pi(q, T)-\Pi(q, 0)=-\frac{g^2}{8\pi^2}\int dk_x \left(48\zeta(3)T^{3-\epsilon}\frac{2\omega_q^2-\tilde{q}^2}{(\tilde{q}^2+\omega_q^2)^2}\right).
\end{equation}
We evaluate $\Pi(q,0)$ using dimensional regularization at finite $v$ to get, to leading order in $\epsilon$:
\begin{align}
&\Pi(q, 0)=-4g^2\int\frac{d^2\mathbf{k}}{(2\pi)^2}\frac{d^{2-\epsilon}\mathbf{K}}{(2\pi)^{2-\epsilon}}\mathrm{Tr}\left[i\gamma_{d-1}G_n(\mathbf{K},\mathbf{k})i\gamma_{d-1}G_{\bar{n}}(\mathbf{K}+\mathbf{Q},\mathbf{k}+\mathbf{q})\right] \nonumber \\
&=-\frac{4g^2}{v}\int\frac{d\varepsilon_n(\mathbf{k})d\varepsilon_{\bar{n}}(\mathbf{k})}{(2\pi)^2}\frac{d^{2-\epsilon}\mathbf{K}}{(2\pi)^{2-\epsilon}}\frac{-\mathbf{K}\cdot(\mathbf{K}+\mathbf{Q})+\varepsilon_n(\mathbf{k})\varepsilon_{\bar{n}}(\mathbf{k}+\mathbf{q})}{[\mathbf{K}^2+\varepsilon_n(\mathbf{k})^2][(\mathbf{K}+\mathbf{Q})^2+\varepsilon_{\bar{n}}(\mathbf{k}+\mathbf{q})^2]} \nonumber \\
&=-\frac{g^2}{v}\int\frac{d^{2-\epsilon}\mathbf{K}}{(2\pi)^{2-\epsilon}}\frac{-\mathbf{K}\cdot(\mathbf{K}+\mathbf{Q})}{[\mathbf{K}^2]^{1/2}[(\mathbf{K}+\mathbf{Q})^2]^{1/2}} \nonumber \\
&= -\frac{g^2}{v}\frac{\mathbf{Q}^{2-\epsilon}}{8\pi\epsilon}=-\frac{g^2}{v}\frac{\mathbf{(\bar{Q}}^2+\omega_q^2)^{1-\epsilon/2}}{8\pi\epsilon},
\end{align}
where the last integral was performed using Feynman parameterization. Inserting the expressions for $\Pi(q,T)-\Pi(q,0)$ and $\Pi(q,0)$ into Eq.~(\ref{eq:ffb0}), we get, using dimensional regularization for the $q$ momentum integrals,
\begin{align}
&F_{fb}=F_{fb}^{(1)}+F_{fb}^{(2)}, \nonumber \\
&F_{fb}^{(1)}=(N_c^2-1)\int\frac{d^2\mathbf{q}d^{1-\epsilon}\mathbf{\bar{Q}}}{(2\pi)^{3-\epsilon}}T\sum_{\omega_q}\frac{\Pi(q,T)-\Pi(q,0)}{\mathbf{\bar{Q}}^2+c^2|\mathbf{q}|^2+\omega_q^2} \nonumber \\
&=(1-N_c^2)T^{3-\epsilon}\frac{6 \zeta(3) g^2}{\pi c}\int dk_x\int\frac{dq_yd^{1-\epsilon}\mathbf{\bar{Q}}}{(2\pi)^{3-\epsilon}}T\sum_{\omega_q}\frac{2\omega_q^2-\tilde{q}^2}{(\tilde{q}^2+\omega_q^2)^2\sqrt{\mathbf{\bar{Q}}^2+\omega_q^2}}
\end{align}
Where we integrated out $q_x$ and then sent $c\rightarrow0$ in the last step of the above. Doing the remaining integrals first over $q_y$ and then $\bar{\mathbf{Q}}$, we get, to leading order in $\epsilon$
\begin{equation}
F_{fb}^{(1)}=\frac{3\zeta(3)g^2(N_c^2-1) T^{3-\epsilon}}{16c\pi^2} \int dk_x T\sum_{\omega_q} \frac{1}{|\omega_q|^{1+\epsilon}} =\int dk_x\frac{3\zeta(3)g^2(N_c^2-1) T^{3-2\epsilon}}{16c\pi^3\epsilon}.
\label{eq:ffren}
\end{equation}
The other part gives
\begin{align}
&F_{fb}^{(2)}=(N_c^2-1)\int\frac{d^2\mathbf{q}d^{1-\epsilon}\mathbf{\bar{Q}}}{(2\pi)^{3-\epsilon}}T\sum_{\omega_q}\frac{\Pi(q,0)}{\mathbf{\bar{Q}}^2+c^2|\mathbf{q}|^2+\omega_q^2} =(1-N_c^2)\frac{g^2}{8\pi v\epsilon}\int\frac{d^2\mathbf{q}d^{1-\epsilon}\mathbf{\bar{Q}}}{(2\pi)^{3-\epsilon}}T\sum_{\omega_q}\frac{(\mathbf{\bar{Q}}^2+\omega_q^2)^{1-\epsilon/2}}{\mathbf{\bar{Q}}^2+c^2|\mathbf{q}|^2+\omega_q^2}. 
\end{align}
Integrating first over $\mathbf{\bar{Q}}$ and then over $\mathbf{q}$ using the dimensional regularization, we get, to leading order in $\epsilon$
\begin{equation}
F_{fb}^{(2)}=(N_c^2-1)\frac{g^2\pi}{6v c^2\epsilon}T\sum_{\omega_q}\frac{1}{|\omega_q|^{-(3-2\epsilon)}}=(N_c^2-1)\frac{g^2\pi}{360v c^2\epsilon}T^{4-2\epsilon}.
\end{equation}

\section{Finite $v$ and $c$}
\label{app:self-one-loop-vcn0}
In this appendix, we describe the breakdown of the results derived in the previous appendices when we do not have $v,c\rightarrow0$. We illustrate this by first computing the self energy correction to $\langle J_y J_y\rangle$ for finite $v$ and $c$; similar problems occur in the computations of the fermion free energy. The fermion self energy for $v,c\neq0$ is given by~\cite{sur15}
\begin{align}
&\Sigma_1(\mathbf{K},\mathbf{k})=-i\frac{\pi^{2-\epsilon/2}\Gamma(\epsilon/2)}{(2\pi)^{4-\epsilon}}\frac{g^2\mu^\epsilon}{cN_f}\int_0^1 dx \sum_{j=1}^{N_c^2-1}(\tau^j\tau^j)
\frac{\mathbf{\Gamma}\cdot\mathbf{K}-\gamma_{d-1}\frac{c^2\varepsilon_{3}(\mathbf{k})}{c^2+x(1+v^2-c^2)}}{\left[\mathbf{K}^2+\frac{c^2\varepsilon_{3}^2(\mathbf{k})}{c^2+x(1+v^2-c^2)}\right]^{\epsilon/2}}\frac{x^{-\epsilon/2}(1-x)^{1/2-\epsilon/2}}{(c^2+x(1+v^2-c^2))^{1/2}},
\label{eq:fse1lvcn0}
\end{align}
We can ignore the term with the prefactor of $c^2$ in the numerator of the integrand in  Eq.~(\ref{eq:fse1lvcn0}); since $v\neq0$ $\varepsilon_{1}(\mathbf{k})$ and $\varepsilon_{3}(\mathbf{k})$ can be taken to be independent variables of integration over $\mathbf{k}$ space via the coordinate transformation $d^2\mathbf{k}\rightarrow d\varepsilon_{1}d\varepsilon_{3}/(2v)$. This term then only produces contributions to $\langle J_y J_y \rangle_{\rm SE}$ that are odd in $\varepsilon_{3}$ and hence vanish under integration over $\varepsilon_{3}$.
Thus dropping this term, we have 
\begin{align}
&\langle J_y J_y \rangle_{\mathrm{SE}}(\omega)= \frac{16(1-N_c^2) \pi^{2-\epsilon/2}\Gamma(\epsilon/2)g^2\mu^\epsilon}{(2\pi)^{8-2\epsilon}c}\int_0^1 dx \frac{x^{-\epsilon/2}(1-x)^{1/2-\epsilon/2}(1+v^2)}{(c^2+x(1+v^2-c^2))^{1/2}} \times \nonumber \\
&\times \int \frac{d\varepsilon_1 d\varepsilon_3}{2v} d^{2-\epsilon}\mathbf{K} \frac{\mathbf{K}^4+\mathbf{W}\cdot\mathbf{K}~\mathbf{K}^2-\varepsilon_1^2(3\mathbf{K}^2+\mathbf{K}\cdot\mathbf{W})}{(\mathbf{K}^2+\varepsilon_1^2)^2((\mathbf{K}+\mathbf{W})^2+\varepsilon_1^2)\left[\mathbf{K}^2+\frac{c^2\varepsilon_3^2}{c^2+x(1+v^2-c^2)}\right]^{\frac{\epsilon}{2}}}
\label{eq:complicated1}
\end{align}
evaluating this as in Appendix~\ref{app:self-one-loop} gives the singular contribution
\begin{align}
&\langle J_y J_y \rangle_{\mathrm{SE}}(\omega)\approx \nonumber \\
&\int \frac{d\varepsilon_3}{2v} (N_c^2-1) \frac{g^2\mu^\epsilon}{c}\omega^{1-2\epsilon}\int_0^1 dx \kappa_1\left(\frac{c^2\varepsilon_3^2}{\omega^2(c^2+x(1+v^2-c^2))},\epsilon\right)\frac{(1-x)^{1/2}(1+v^2)}{(c^2+x(1+v^2-c^2))^{1/2}}\left(\frac{1}{64 \pi ^3 \epsilon}\right),
\label{eq:complicated2}
\end{align}
where the crossover function $\kappa_1(x,\epsilon)\approx (1+x)^{-\epsilon/2}$ for $x \ll 1$. 

The singular part of the 1-loop current vertex at finite $v$ and $c$ is most easily derived from the Ward identity;
\begin{align}
&\Xi_3(\mathbf{K},\mathbf{k},0)\Bigg|_{\rm pole}=-\frac{d \Sigma_1(\mathbf{K},\mathbf{k})}{d k_y}\Bigg|_{\rm pole}= \nonumber \\
&i\gamma_{d-1}\frac{\pi^{2-\epsilon/2}\Gamma(\epsilon/2)}{(2\pi)^{4-\epsilon}}\frac{g^2c\mu^\epsilon}{N_f}\sum_{j=1}^{N_c^2-1}(\tau^j\tau^j)\int_0^1 dx 
\left[\mathbf{K}^2+\frac{c^2\varepsilon_{3}^2(\mathbf{k})}{c^2+x(1+v^2-c^2)}\right]^{-\epsilon/2}\frac{x^{-\epsilon/2}(1-x)^{1/2-\epsilon/2}}{(c^2+x(1+v^2-c^2))^{3/2}}.
\label{eq:vertpolevcn0}
\end{align}
Inserting this into Eq.~(\ref{eq:two-loop-vert}) gives, for the singular part of the two-loop vertex correction to $\langle J_y J_y \rangle$, via a computation very similar to that for the self energy correction (an additional prefactor of 2 has to be inserted to account for both the poles associated with vertex corrections to each of the two current vertices in the graph),
\begin{align}
&\langle J_y J_y \rangle_{\mathrm{vert}}(\omega) \approx -\frac{g^2c\mu^\epsilon(N_c^2-1)}{8\pi^6\epsilon}\int \frac{d\varepsilon_1d\varepsilon_3}{2v}d^{2-\epsilon}\mathbf{K}\int_0^1 dx \frac{(1-x)^{1/2}(1-v^2)}{(c^2+x(1+v^2-c^2))^{3/2}} \times \nonumber \\
&\Bigg[\frac{-\mathbf{K}\cdot(\mathbf{K}+\mathbf{W})+\varepsilon_1^2}{(\mathbf{K}^2+\varepsilon_1^2)((\mathbf{K}+\mathbf{W})^2+\varepsilon_1^2)\left[\mathbf{K}^2+\frac{c^2\varepsilon_{3}^2}{c^2+x(1+v^2-c^2)}\right]^{\epsilon/2}}\Bigg] \nonumber \\
&\approx - \int\frac{d\varepsilon_3}{2v}(N_c^2-1) g^2c\mu^\epsilon\omega^{1-2\epsilon}\int_0^1 dx \kappa_2\left(\frac{c^2\varepsilon_3^2}{\omega^2(c^2+x(1+v^2-c^2))},\epsilon\right) \frac{(1-x)^{1/2}(1-v^2)}{(c^2+x(1+v^2-c^2))^{3/2}}\left(\frac{1}{32\pi^3\epsilon}\right),
\label{eq:two-loop-vertvcn0}
\end{align}
where again the crossover function $\kappa_2(x,\epsilon)\approx (1+x)^{-\epsilon/2}$ for $x \ll 1$. 

\section{Boltzmann Equation Computations}
\label{app:Boltzmann}

\subsection{Collisionless conductivity in $d=3-\epsilon$}
We can diagonalize the Hamiltonian corresponding to the free fermion part of Eq.~(\ref{eq:embedded_action}) as 
\begin{align}
&\mathcal{H}_f = \sum_{n=1}^{4} \sum_{\sigma=1}^{N_c} \sum_{j=1}^{N_f} \int \frac{d^2\mathbf{k}d^{1-\epsilon}\mathbf{\bar{K}}}{(2\pi)^{3-\epsilon}} \bar{\Psi}_{n,\sigma,j}(\mathbf{k},\mathbf{\bar{K}}) \left[ i\bar{\mathbf{\Gamma}} \cdot \bar{\mathbf{K}} + i \gamma_{d-1} \varepsilon_n(\mathbf{k})\right] \Psi_{n,\sigma,j}(\mathbf{k},\mathbf{\bar{K}}) \nonumber \\
& = \sum_{n=1}^{4} \sum_{\sigma=1}^{N_c} \sum_{j=1}^{N_f} \sum_{m=\pm} \int \frac{d^2\mathbf{k}d^{1-\epsilon}\mathbf{\bar{K}}}{(2\pi)^{3-\epsilon}} \lambda^{\dagger}_{n,\sigma,j,m}(\mathbf{k},\mathbf{\bar{K}})\xi_{n,m}(\mathbf{k},\mathbf{\bar{K}}) \lambda_{n,\sigma,j,m}(\mathbf{k},\mathbf{\bar{K}})
\end{align}
with the particle-hole symmetric dispersions $\xi_{n,m}(\mathbf{k},\mathbf{\bar{K}})=m\left(\bar{\mathbf{K}}^2+\varepsilon^2_n(\mathbf{k})\right)^{1/2}$. The physical current density becomes
\begin{equation}
\mathbf{J}= \sum_{n=1}^{4} \sum_{\sigma=1}^{N_c} \sum_{j=1}^{N_f} \sum_{m=\pm} \int \frac{d^2\mathbf{k}d^{1-\epsilon}\mathbf{\bar{K}}}{(2\pi)^{3-\epsilon}} \left(\mathbf{v}_n \frac{m\varepsilon_n(\mathbf{k})}{\sqrt{\bar{\mathbf{K}}^2+\varepsilon_n(\mathbf{k})^2}} \lambda^{\dagger}_{n,\sigma,j,m}(\mathbf{k},\mathbf{\bar{K}}) \lambda_{n,\sigma,j,m}(\mathbf{k},\mathbf{\bar{K}})\right)+\mathbf{J}_2,
\end{equation}
where $\varepsilon_n(\mathbf{k})=\mathbf{v}_n\cdot\mathbf{k}$ and $\mathbf{J}_2$ contains particle-hole terms $\lambda^{\dagger}_+\lambda_-,~\lambda^{\dagger}_-\lambda_+$ that are unimportant for transport in the low frequency regime of $\omega\ll T$~\cite{fritz08,fritz11}. Defining the distribution functions
\begin{equation}
f_{n,m}(\mathbf{k},\mathbf{\bar{K}},t)=\langle\lambda^{\dagger}_{n,\sigma,j,m}(\mathbf{k},\mathbf{\bar{K}},t) \lambda_{n,\sigma,j,m}(\mathbf{k},\mathbf{\bar{K}},t)\rangle,
\end{equation}
we have the collisionless kinetic equation in the presence of an applied electric field
\begin{equation}
\left(\frac{\partial}{\partial t}+m\mathbf{E}\cdot\frac{\partial}{\partial\mathbf{k}}\right)f_{n,m}(\mathbf{k},\mathbf{\bar{K}},t)=0,
\end{equation}
with the frequency-domain solution to linear order in $\mathbf{E}$
\begin{equation}
f_{n,m}(\mathbf{k},\mathbf{\bar{K}},\omega)=2\pi\delta(\omega)n_f(\xi_{n,m}(\mathbf{k},\mathbf{\bar{K}}))+\mathbf{v}_n\cdot\mathbf{E}(\omega)\frac{m\varepsilon_n(\mathbf{k})}{\sqrt{\bar{\mathbf{K}}^2+\varepsilon_n(\mathbf{k})^2}}\frac{1/T}{-i\omega+0^+}n_f(\xi_{n,m}(\mathbf{k},\mathbf{\bar{K}}))(1-n_f(\xi_{n,m}(\mathbf{k},\mathbf{\bar{K}}))).
\end{equation}
Inserting this into the expression for $\mathbf{J}$, we obtain the collisionless conductivity
\begin{align}
&\sigma_{xx}(\omega)=\frac{\delta J_x(\omega)}{\delta E_x(\omega)}=4N_cN_f \frac{(1+v^2)/T}{-i\omega+0^+} \int \frac{d^2\mathbf{k}d^{1-\epsilon}\mathbf{\bar{K}}}{(2\pi)^{3-\epsilon}} \frac{\varepsilon_n^2(\mathbf{k})}{\mathbf{\bar{K}}^2+\varepsilon_n^2(\mathbf{k})}n_f(\xi_{n,+}(\mathbf{k},\mathbf{\bar{K}}))(1-n_f(\xi_{n,+}(\mathbf{k},\mathbf{\bar{K}}))), \nonumber \\
&\mathrm{Re}[\sigma_{xx}(\omega)]=2N_cN_f\int dk_\parallel\sqrt{1+v^2}\frac{\delta(\omega)}{T}\int\frac{d\varepsilon_nd^{1-\epsilon}\mathbf{\bar{K}}}{(2\pi)^{2-\epsilon}}\frac{\varepsilon_n^2}{\mathbf{\bar{K}}^2+\varepsilon_n^2}n_f\left(\sqrt{\mathbf{\bar{K}}^2+\varepsilon_n^2}\right)\left(1-n_f\left(\sqrt{\mathbf{\bar{K}}^2+\varepsilon_n^2}\right)\right) \nonumber \\
&=2N_cN_f\sqrt{1+v^2}\int dk_\parallel\delta(\omega)T^{1-\epsilon}\frac{\pi^{1-\epsilon/2}(1-2^\epsilon)\Gamma(2-\epsilon)\zeta(1-\epsilon)}{(2\pi)^{2-\epsilon}\Gamma(2-\epsilon/2)}=\mathrm{Re}[\sigma_{yy}(\omega)].
\end{align}

\subsection{Derviation of the fermion collision integral}

We derive the following expressions for the different components of the fermion self energies in the Keldysh formalism. For a fermion at hot spot given by $(\ell,+)$, we get for the first diagram for $\Sigma^{R}_f$ in Fig.~\ref{fig:keldyshgraphs}
\begin{align}
\Sigma^{R \ell +(1)}_{f,\sigma\sigma'}(x,x')
&=
i g^2 \tau^a_{\sigma\rho} \tau^a_{\rho\sigma'}
D^K_0 (x, x') G^{R \ell -}_0 (x,x') =3i\delta_{\sigma\sigma'}g^2 D^K_0 (x, x') G^{R\ell -}_0 (x,x'). 
\end{align}
We use that for products of Wigner transforms
\begin{align}
D_0^{K}(x,x') G_0^{R\ell-}(x,x') \rightarrow \sum_{q} D_0^{K}(x,p-q) G_0^{R\ell-}(x,q),
\end{align}
and plug in the representation of the Keldysh propagator in terms of the distribution function to get
\begin{align}
\Sigma^{R\ell+(1)}_{f,\sigma\sigma'}(x,p)
= 3 \delta_{\sigma\sigma'}g^2 
\sum_{q}
F_b(x,p-q) 
\,
i \left[
D^R_0(x,p-q) - D_0^A(x,p-q)
\right]
G_0^{R\ell-}(x,q).
\end{align}
For the collision integral on the right hand side of (\ref{eq:fke}) we need twice the imaginary part of this expression.
Using
\begin{align}
2 {\rm Im }[G_0^{R\ell-}(x,q)] = \frac{1}{i} 
\left[
G_0^{R\ell-}(x,q) - G_0^{A\ell-}(x,q)
\right],
\end{align}
we get,
\begin{align}
&2 {\rm Im} \left[ \Sigma^{R\ell+(1)}_{f,\sigma\sigma'}(x,p)\right]
\nonumber\\
&=
3 \delta_{\sigma\sigma'} g^2 
\sum_q F_b(x,p-q)
\,
i \left[
D^R_0(x,p-q) - D_0^A(x,p-q)
\right]
\frac{1}{i} 
\left[
G_0^{R\ell-}(x,q) - G_0^{A\ell-}(x,q)
\right]
\nonumber\\
&=
-3
\delta_{\sigma\sigma'} g^2 
\int d^2 \mathbf{q} \int \frac{ d\omega}{2\pi}
\frac{1}{4 \omega_{\mathbf{p}-\mathbf{q}}}
\left(
\delta(e^{+}_{\ell}(\mathbf{p}) - \omega - \omega_{\mathbf{p}-\mathbf{q}})
-
\delta(e^{+}_{\ell}(\mathbf{p}) - \omega + \omega_{\mathbf{p}-\mathbf{q}})
\right)\times \nonumber \\
&\times\delta(\omega - e^{-}_{\ell}(\mathbf{q}))
F_b(t,\mathbf{p}-\mathbf{q},e_\ell^+(\mathbf{p})-\omega) 
\nonumber\\
&
=
- 3
\delta_{\sigma\sigma'} g^2 
\int \frac{ d^2 \mathbf{q}} {2 \pi} 
\frac{1}{4 \omega_{\mathbf{p}-\mathbf{q}}}
\Big(
\delta(e^{+}_{\ell}(\mathbf{p}) -e^{-}_{\ell}(\mathbf{q}) - \omega_{\mathbf{p}-\mathbf{q}})F_b(t,\mathbf{p}-\mathbf{q},\omega_{\mathbf{p}-\mathbf{q}}) - \nonumber \\
&-
\delta(e^{+}_{\ell}(\mathbf{p}) - e^{-}_{\ell}(\mathbf{q}) + \omega_{\mathbf{p}-\mathbf{q}})F_b(t,\mathbf{p}-\mathbf{q},-\omega_{\mathbf{p}-\mathbf{q}}) 
\Big) \nonumber \\
&=2 \delta_{\sigma\sigma'} {\rm Im} \left[ \Sigma^{R\ell+(1)}_{f}(t,\mathbf{p},e_\ell^+(\mathbf{p}))\right],
\label{eq:Sigmaf_1}
\end{align}
where we have used spatial translational invariance and also have kept the external fermion on shell. Likewise, for the second diagram for $\Sigma^{R}_f$ in Fig.~\ref{fig:keldyshgraphs} contributing to $\Sigma^{R\ell+}_{f,\sigma\sigma^\prime}$ we have
\begin{align}
&2\mathrm{Im}[\Sigma^{R\ell+(2)}_{f,\sigma\sigma^\prime}(x,p)] \nonumber \\
&=3\delta_{\sigma\sigma^\prime}g^2\sum_q F_f(x,q)[G_0^{R\ell-}(x,q)-G_0^{A\ell-}(x,q)][D_0^R(x,p-q)-D_0^A(x,p-q)] \nonumber \\
&=-3\delta_{\sigma\sigma^\prime}g^2\int\frac{d^2\mathbf{q}}{2\pi}\frac{1}{4\omega_{\mathbf{p}-\mathbf{q}}}\left(\delta(e^{+}_{\ell}(\mathbf{p}) - e^{-}_{\ell}(\mathbf{q}) - \omega_{\mathbf{p}-\mathbf{q}})-\delta(e^{+}_{\ell}(\mathbf{p}) - e^{-}_{\ell}(\mathbf{q}) + \omega_{\mathbf{p}-\mathbf{q}})\right)F_f^{\ell-}(t,\mathbf{q}) \nonumber \\
&=2 \delta_{\sigma\sigma'} {\rm Im} \left[ \Sigma^{R\ell+(2)}_{f}(t,\mathbf{p},e_\ell^+(\mathbf{p}))\right].
\end{align}
For the diagrams in Fig.~\ref{fig:keldyshgraphs} contributing to $\Sigma^{K\ell+}_{f,\sigma\sigma^\prime}$, the first gives
\begin{align}
&i\Sigma^{K\ell+(1)}_{f,\sigma\sigma^\prime}(x,p) \nonumber \\
&=-3\delta_{\sigma\sigma^\prime}g^2\sum_q F_f(x,q)F_b(x,p-q)[G_0^{R\ell-}(x,q)-G_0^{A\ell-}(x,q)][D_0^R(x,p-q)-D_0^A(x,p-q)] \nonumber \\
&=3\delta_{\sigma\sigma^\prime}g^2\int\frac{d^2\mathbf{q}}{2\pi}\frac{1}{4\omega_{\mathbf{p}-\mathbf{q}}}\Bigg(\delta(e^{+}_{\ell}(\mathbf{p}) - e^{-}_{\ell}(\mathbf{q}) - \omega_{\mathbf{p}-\mathbf{q}})F_f^{\ell-}(t,\mathbf{q})F_b(t,\mathbf{p}-\mathbf{q},\omega_{\mathbf{p}-\mathbf{q}}) - \nonumber \\
&- \delta(e^{+}_{\ell}(\mathbf{p}) - e^{-}_{\ell}(\mathbf{q}) + \omega_{\mathbf{p}-\mathbf{q}})F_f^{\ell-}(t,\mathbf{q})F_b(t,\mathbf{p}-\mathbf{q},-\omega_{\mathbf{p}-\mathbf{q}}) \Bigg) \nonumber \\
&=2i \delta_{\sigma\sigma'} \Sigma^{K\ell+(1)}_{f}(t,\mathbf{p},e_\ell^+(\mathbf{p})).
\end{align}
The second and third combined yield
\begin{align}
&i\Sigma^{K\ell+(2+3)}_{f,\sigma\sigma^\prime}(x,p) \nonumber \\
&=-3\delta_{\sigma\sigma^\prime}g^2\sum_q [G_0^{R\ell-}(x,q)D_0^R(x,p-q)+G_0^{A\ell-}D_0^A(x,p-q)] \nonumber \\
&=3\delta_{\sigma\sigma^\prime}\int\frac{d^2\mathbf{q}}{4\pi^2}\int\frac{d\omega}{2\pi}\frac{g^2}{4\omega_{\mathbf{p}-\mathbf{q}}}\Bigg[\frac{1}{\omega-e^{-}_{\ell}(\mathbf{q})+i0^+}\Bigg(\frac{1}{e^{+}_{\ell}(\mathbf{p})-\omega+\omega_{p-q}+i0^+}-\frac{1}{e^{+}_{\ell}(\mathbf{p})-\omega-\omega_{p-q}+i0^+}\Bigg)+\mathrm{c.c.}\Bigg] \nonumber \\
&=3\delta_{\sigma\sigma^\prime}g^2\int\frac{d^2\mathbf{q}}{2\pi}\frac{1}{4\omega_{\mathbf{p}-\mathbf{q}}}\left(\delta(e^{+}_{\ell}(\mathbf{p}) - e^{-}_{\ell}(\mathbf{q}) - \omega_{\mathbf{p}-\mathbf{q}})-\delta(e^{+}_{\ell}(\mathbf{p}) - e^{-}_{\ell}(\mathbf{q}) + \omega_{\mathbf{p}-\mathbf{q}})\right) \nonumber \\
&=2i \delta_{\sigma\sigma'} \Sigma^{K\ell+(2+3)}_{f}(t,\mathbf{p},e_\ell^+(\mathbf{p})).
\end{align}
Combining the above expressions gives the collision integral for fermions of any spin at the hot spot given by ($\ell$, +)
\begin{align}
&I^{\rm coll}_{f\ell+}[F_f,F_b](t,\mathbf{p})= \nonumber \\
&=3g^2\int d^2\mathbf{q}\frac{1}{4\omega_{\mathbf{p}-\mathbf{q}}}\Bigg(\delta(e^{+}_{\ell}(\mathbf{p}) - e^{-}_{\ell}(\mathbf{q}) - \omega_{\mathbf{p}-\mathbf{q}})\Big[1+F_f^{\ell-}(t,\mathbf{q})F_b(t,\mathbf{p}-\mathbf{q},\omega_{\mathbf{p}-\mathbf{q}})-F_f^{\ell-}(t,\mathbf{q})F_f^{\ell+}(t,\mathbf{p})- \nonumber \\
& - F_f^{\ell+}(t,\mathbf{p})F_b(t,\mathbf{p}-\mathbf{q},\omega_{\mathbf{p}-\mathbf{q}})\Big] -\delta(e^{+}_{\ell}(\mathbf{p}) - e^{-}_{\ell}(\mathbf{q}) + \omega_{\mathbf{p}-\mathbf{q}})\Big[1+F_f^{\ell-}(t,\mathbf{q})F_b(t,\mathbf{p}-\mathbf{q},-\omega_{\mathbf{p}-\mathbf{q}})- \nonumber \\
&-F_f^{\ell-}(t,\mathbf{q})F_f^{\ell+}(t,\mathbf{p})-F_f^{\ell+}(t,\mathbf{p})F_b(t,\mathbf{p}-\mathbf{q},-\omega_{\mathbf{p}-\mathbf{q}})\Big]\Bigg). 
\end{align}
The collision integral for the hot spot given by ($\ell$, -) is given by simply interchanging $+\leftrightarrow-$ in the above. With the alternate parameterization (\ref{eq:distribother}) of the distribution functions, and relabeling of momenta $\mathbf{q}\leftrightarrow\mathbf{p}-\mathbf{q}$ we may rewrite this for any hot spot as
\begin{align}
&I^{\rm coll}_{f\ell\pm}[f_f,f_b](t,\mathbf{p})= \nonumber \\
&=3g^2\int\frac{d^2\mathbf{q}}{2\pi}\frac{1}{\omega_{\mathbf{q}}}\Bigg(\delta(e^{\pm}_{\ell}(\mathbf{p}) - e^{\mp}_{\ell}(\mathbf{p}-\mathbf{q}) - \omega_{\mathbf{q}})\Big[f_f^{\ell\pm}(t,\mathbf{p})(1-f_f^{\ell\mp}(t,\mathbf{p}-\mathbf{q}))-f_f^{\ell\mp}(t,\mathbf{p}-\mathbf{q})f_b(t,\mathbf{q},\omega_{\mathbf{q}})+ \nonumber \\
&+f_f^{\ell\pm}(t,\mathbf{p})f_b(t,\mathbf{q},\omega_{\mathbf{q}})\Big] -\delta(e^{\pm}_{\ell}(\mathbf{p}) -e^{\mp}_{\ell}(\mathbf{p}-\mathbf{q}) + \omega_{\mathbf{q}})\Big[f_f^{\ell\pm}(t,\mathbf{p})(1-f_f^{\ell-}(t,\mathbf{p}-\mathbf{q}))-\nonumber \\
&-f_f^{\ell\mp}(t,\mathbf{p}-\mathbf{q})f_b(t,\mathbf{q},-\omega_{\mathbf{q}})+f_f^{\ell\pm}(t,\mathbf{p})f_b(t,\mathbf{q},-\omega_{\mathbf{q}})\Big]\Bigg).
\end{align}

\subsection{Derivation of the boson collision integral}

We begin with the retarded component of the boson self energy in the Keldysh formalism. The sum of the two diagrams for this component in Fig.~\ref{fig:keldyshgraphs} gives
\begin{align}
\Sigma_{b}^R
(x,x')
=
-i g^2\sum_{\ell} 
\left[
G_0^{K\ell+}(x,x') G_0^{A\ell-}(x',x) 
+ 
G_0^{R\ell-}(x,x')G_0^{K\ell+}(x',x)  
+
( + \leftrightarrow -)
\right].
\end{align}
Wigner transforming this gives
\begin{align}
&2 {\rm Im} [\Sigma^R_{b}(x,q)]
=
-g^2\sum_{\ell}
\sum_k\Big[
F_f(x,k+q) 
\left(
G_0^{R\ell+}(x,k+q) - G_0^{A\ell+}(x,k+q) 
\right)
\left(
G_0^{A\ell-}(x,k) - G_0^{R\ell-}(x,k) 
\right)
\nonumber \\
&
+
F_f(x,k)
\left(
G_0^{R\ell+}(x,k) - G_0^{A\ell+}(x,k) 
\right)
\left(
G_0^{R\ell-}(x,k+q) - G_0^{A\ell-}(x,k+q) 
\right)
+
( + \leftrightarrow -)
\Big]\nonumber\\
&
=
-g^2 \sum_{\ell}
\int \frac{d^2 \mathbf{k}}{2 \pi}
\Big[
\delta\left(
e^-_\ell(\mathbf{k})
+ 
\omega_{\mathbf{q}}
-
e^+_\ell(\mathbf{k}+\mathbf{q})
\right)
F_f^{\ell+}(t,\mathbf{k}+\mathbf{q})
- \nonumber \\
&-
\delta\left(
e^+_\ell(\mathbf{k})
+ 
\omega_{\mathbf{q}}
-
e^-_\ell(\mathbf{k}+\mathbf{q})
\right)
F_f^{\ell+}(t,\mathbf{k})
+
( + \leftrightarrow -)
\Big] \nonumber \\
&=2{\rm Im}[\Sigma_b^R(t,\mathbf{q},\omega_\mathbf{q})],
\end{align}
where we have used spatial translational invariance and also have kept the external boson on shell. For the Keldysh component of the boson self energy, the second diagram in Fig.~\ref{fig:keldyshgraphs} gives
\begin{align}
&i\Sigma^{K(2)}_{b}(x,q) \nonumber \\
&= 
g^2 \sum_{\ell} \sum_k
\Big[
F_f(x,k+q)
\left(
G_0^{R\ell+}(x,k+q) - G_0^{A\ell+}(x,k+q)
\right) F_f(x,k)
\left(
G_0^{R\ell-}(x,k) - G_0^{A\ell-}(x,k)
\right)
+ ( + \leftrightarrow -)
\Big]
\nonumber \\
&=
- i g^2\sum_{\ell}
\int \frac{d^2 \mathbf{k}}{2\pi}
\left[
\delta\left(\omega_{\bf{q}} + e^-_\ell(\mathbf{k}) -  e^+_\ell(\mathbf{k}+\mathbf{q}) \right)
F_f^{\ell+}(t,\mathbf{k}+\mathbf{q}) F_f^{\ell-}(t,\mathbf{k})
+
( + \leftrightarrow -)
\right] \nonumber \\
&=2i\Sigma_b^{K(2)}(t,\mathbf{q},\omega_\mathbf{q}).
\end{align}
The first and third diagrams for the Keldysh component when combined give
\begin{align}
i\Sigma^{K(1+3)}_{b}(x,q)
&= 
g^2\sum_{\ell} \sum_k
\left[
\left(
G_0^{R\ell+}(x,k+q)G_0^{A\ell+}(x,k)
\right)
+
\left(
G_0^{A\ell+}(x,k+q) G_0^{R\ell-}(x,k)
\right)
+ ( + \leftrightarrow -)
\right]
\nonumber\\
&
=
g^2\sum_{\ell} 
\int \frac{d^2 \mathbf{k}}{4 \pi^2}\int\frac{d\omega}{2\pi}
\left[\frac{1}{\omega_{\mathbf{q}}+\omega-e_{\ell}^+(\mathbf{k}+\mathbf{q})+i0^+}\frac{1}{\omega-e_{\ell}^-(\mathbf{k})-i0^+}+\mathrm{c.c.}
+
( + \leftrightarrow -)
\right]\;
\nonumber \\
&
=
g^2 \sum_{\ell}
\int \frac{d^2 \mathbf{k}}{2\pi}
\left[\delta\left(\omega_{\mathbf{q}}+e_{\ell}^-(\mathbf{k})-e_{\ell}^+(\mathbf{k}+\mathbf{q})\right)
+
( + \leftrightarrow -)
\right] \nonumber \\
&=2i\Sigma_b^{K(1+3)}(t,\mathbf{q},\omega_\mathbf{q}).
\end{align}
Thus we obtain the boson collision integral:
\begin{align}
&I^{\rm coll}_{b}[F_f,F_b](t,\mathbf{q})=-g^2
\sum_{\ell}
\int\frac{d^2\mathbf{k}}{2\pi}\Big[\delta\left(e^-_\ell(\mathbf{k})+ \omega_{\mathbf{q}}-e^+_\ell(\mathbf{k}+\mathbf{q})\right)
\nonumber\\
&\Big(-1-F_{f}^{\ell-}(t,\mathbf{k})F_b(t,\mathbf{q},\omega_{\mathbf{q}})+F_{f}^{\ell+}(t,\mathbf{k}+\mathbf{q})F_{f}^{\ell-}(t,\mathbf{k})
+ F_{f}^{\ell+}(t,\mathbf{k}+\mathbf{q})F_b(t,\mathbf{q},\omega_{\mathbf{q}})\Big)+( + \leftrightarrow -)\Big], \nonumber \\
\end{align}
or
\begin{align}
&I^{\rm coll}_{b}[f_f,f_b](t,\mathbf{q})=4g^2\sum_{\ell} \int\frac{d^2\mathbf{k}}{2\pi}\Big[\delta\left(e^-_\ell(\mathbf{k})+ \omega_{\mathbf{q}}-e^+_\ell(\mathbf{k}+\mathbf{q})\right)\Big(f_{f}^{\ell+}(t,\mathbf{k}+\mathbf{q})(1-f_{f}^{\ell-}(t,\mathbf{k}))+ \nonumber \\
&+ f_{f}^{\ell+}(t,\mathbf{k}+\mathbf{q})f_b(t,\mathbf{q},\omega_{\mathbf{q}})-f_{f}^{\ell-}(t,\mathbf{k})f_b(t,\mathbf{q},\omega_{\mathbf{q}})\Big) + (+\leftrightarrow-)\Big]. 
\end{align}

\subsection{Solution of the boson kinetic equation}

Inserting the parametrization Eq.~(\ref{eq:fermionpar}) for the fermion $f$ functions into the boson collision integral and also parameterizing $f_b$ in the frequency domain as
\begin{equation}
f_b(\omega,\mathbf{q},\omega_{\mathbf{q}})=2\pi\delta(\omega)n_b(\omega_\mathbf{q})+u(\omega, \mathbf{q},\omega_{\mathbf{q}}),
\end{equation}
where $u$ is linear in $\mathbf{E}$. Changing variables to $p_1^\ell=e_\ell^-(\mathbf{k})=\mathbf{v}_\ell^-\cdot\mathbf{k}$ and $p_2^\ell=e_\ell^+(\mathbf{k})=\mathbf{v}_\ell^+\cdot\mathbf{k}$, the boson collision integral (\ref{eq:bci}) becomes
\begin{align}
&I^{coll}_{b}[F_f,F_b](t,\mathbf{q}) \nonumber \\
&=\frac{g^2}{\pi v}\sum_{\ell} \int dp_1^\ell dp_2^\ell\Big[\delta\left(p_1^\ell+ \omega_{\mathbf{q}}-p_2^\ell-\mathbf{v}_\ell^+\cdot{\mathbf{q}}\right)\Big(f_{\ell}^+(t,\mathbf{k}+\mathbf{q})\left(1-f_{\ell}^-(t,\mathbf{k})\right)+ \nonumber \\
&+f_{\ell}^+(t,\mathbf{k}+\mathbf{q})f_b(t,\mathbf{q},\omega_{\mathbf{q}})-f_{\ell}^-(t,\mathbf{k})f_b(t,\mathbf{q},\omega_{\mathbf{q}})\Big) +  (+\leftrightarrow-,1\leftrightarrow2)\Big]. 
\end{align}
Always integrating out $p_2^\ell$ in this expression, and keeping only terms up to linear order in $\mathbf{E}$, we get, in the frequency domain, using the boson kinetic equation Eq.~(\ref{eq:bke})
\begin{align}
&2(-i\omega+0^+)u(\omega,\mathbf{q},\omega_{\mathbf{q}})= I^{coll}_{b}[F_f,F_b](\omega) \nonumber \\
&=\frac{g^2}{\pi v}\sum_{\ell} \int dp_1^\ell\Big[a(p_1^\ell,\omega_\mathbf{q},\mathbf{v}_\ell^-\cdot\mathbf{q})(2\pi\delta(\omega)n_b(\omega_{\mathbf{q}})+ u(\omega,\mathbf{q},\omega_{\mathbf{q}}))+ \nonumber \\
&+ \mathbf{E}(\omega)\cdot\mathbf{b}^\ell(p_1^\ell,\omega_{\mathbf{q}},\mathbf{v}_\ell^-\cdot\mathbf{q})n_b(\omega_{\mathbf{q}})+ \mathbf{E}(\omega)\cdot\mathbf{b}^\ell_1(p_1^\ell,\omega_{\mathbf{q}},\mathbf{v}_\ell^-\cdot\mathbf{q})- \mathbf{E}(\omega)\cdot\mathbf{d}^\ell(p_1^\ell,\omega_{\mathbf{q}},\mathbf{v}_\ell^-\cdot\mathbf{q})- \nonumber \\
&-c(p_1^\ell,\omega_{\mathbf{q}},\mathbf{v}_\ell^-\cdot\mathbf{q})(2\pi\delta(\omega))+a_1(p_1^\ell,\omega_{\mathbf{q}},\mathbf{v}_\ell^-\cdot\mathbf{q})(2\pi\delta(\omega))\Big], 
\end{align} 
where
\begin{align}
& a(p_1^\ell,\omega_\mathbf{q},\mathbf{v}_\ell^-\cdot\mathbf{q})=(n_f(p_1^\ell+\omega_{\mathbf{q}})-n_f(p_1^\ell))+(n_f(p_1^\ell+\mathbf{v}_\ell^-\cdot\mathbf{q})-n_f(p_1^\ell+\mathbf{v}_\ell^-\cdot\mathbf{q}-\omega_\mathbf{q})), \nonumber \\
&\int dp_1^\ell a(p_1^\ell,\omega_\mathbf{q},\mathbf{v}_\ell^-\cdot\mathbf{q}) = -2\omega_{\mathbf{q}}, \nonumber \\
& \mathbf{b}^\ell(p_1^\ell,\omega_\mathbf{q},\mathbf{v}_\ell^-\cdot\mathbf{q})= \mathbf{v}_\ell^+n_f(p_1^\ell+\omega_{\mathbf{q}})(1-n_f(p_1^\ell+\omega_{\mathbf{q}}))\varphi(p_1^\ell+\omega_{\mathbf{q}})-\mathbf{v}_\ell^-n_f(p_1^\ell)(1-n_f(p_1^\ell))\varphi(p_1^\ell)+ \nonumber \\
&+\mathbf{v}_\ell^-n_f(p_1^\ell+\mathbf{v}_\ell^-\cdot\mathbf{q})(1-n_f(p_1^\ell+\mathbf{v}_\ell^-\cdot\mathbf{q}))\varphi(p_1^\ell+\mathbf{v}_\ell^-\cdot\mathbf{q})- \nonumber \\
&-\mathbf{v}_\ell^+n_f(p_1^\ell+\mathbf{v}_\ell^-\cdot\mathbf{q}-\omega_\mathbf{q})(1-n_f(p_1^\ell+\mathbf{v}_\ell^-\cdot\mathbf{q}-\omega_\mathbf{q}))\varphi(p_1^\ell+\mathbf{v}_\ell^-\cdot\mathbf{q}-\omega_\mathbf{q}), \nonumber \\
&\mathbf{b}_1^\ell(p_1^\ell,\omega_\mathbf{q},\mathbf{v}_\ell^-\cdot\mathbf{q})= \mathbf{v}_\ell^+ n_f(p_1^\ell+\omega_\mathbf{q})(1-n_f(p_1^\ell+\omega_\mathbf{q}))\varphi(p_1^\ell+\omega_\mathbf{q})+  \nonumber \\
&+\mathbf{v}_\ell^- n_f(p_1^\ell+\mathbf{v}_\ell^-\cdot\mathbf{q})(1-n_f(p_1^\ell+\mathbf{v}_\ell^-\cdot\mathbf{q})) \varphi(p_1^\ell+\mathbf{v}_\ell^-\cdot\mathbf{q}), \nonumber \\
&\mathbf{d}^\ell(p_1^\ell,\omega_\mathbf{q},\mathbf{v}_\ell^-\cdot\mathbf{q})=\mathbf{v}_\ell^-n_f(p_1^\ell+\omega_\mathbf{q})n_f(p_1^\ell)(1-n_f(p_1^\ell))\varphi(p_1^\ell)+ \nonumber \\
&+\mathbf{v}_\ell^+n_f(p_1^\ell)n_f(p_1^\ell+\omega_\mathbf{q})(1-n_f(p_1^\ell+\omega_\mathbf{q}))\varphi(p_1^\ell+\omega_\mathbf{q}) \nonumber \\
&+\mathbf{v}_\ell^+n_f(p_1^\ell+\mathbf{v}_\ell^-\cdot\mathbf{q})n_f(p_1^\ell+\mathbf{v}_\ell^-\cdot\mathbf{q}-\omega_\mathbf{q})(1-n_f(p_1^\ell+\mathbf{v}_\ell^-\cdot\mathbf{q}-\omega_\mathbf{q}))\varphi(p_1^\ell+\mathbf{v}_\ell^-\cdot\mathbf{q}-\omega_\mathbf{q}) + \nonumber \\
&+\mathbf{v}_\ell^-n_f(p_1^\ell+\mathbf{v}_\ell^-\cdot\mathbf{q}-\omega_\mathbf{q})n_f(p_1^\ell+\mathbf{v}_\ell^-\cdot\mathbf{q})(1-n_f(p_1^\ell+\mathbf{v}_\ell^-\cdot\mathbf{q}))\varphi(p_1^\ell+\mathbf{v}_\ell^-\cdot\mathbf{q}), \nonumber \\
&c(p_1^\ell,\omega_\mathbf{q},\mathbf{v}_\ell^-\cdot\mathbf{q})=n_f(p_1^\ell+\omega_\mathbf{q})n_f(p_1^\ell)+n_f(p_1^\ell+\mathbf{v}_\ell^-\cdot\mathbf{q})n_f(p_1^\ell+\mathbf{v}_\ell^-\cdot\mathbf{q}-\omega_\mathbf{q}), \nonumber \\
&a_1(p_1^\ell,\omega_\mathbf{q},\mathbf{v}_\ell^-\cdot\mathbf{q})=n_f(p_1^\ell+\omega_\mathbf{q})+n_f(p_1^\ell+\mathbf{v}_\ell^-\cdot\mathbf{q}), \nonumber \\
&\int dp_1^\ell \left[a_1(p_1^\ell,\omega_\mathbf{q},\mathbf{v}_\ell^-\cdot\mathbf{q})-c(p_1^\ell,\omega_\mathbf{q},\mathbf{v}_\ell^-\cdot\mathbf{q})\right]=2\int dp_1^\ell n_f(p_1^\ell+\omega_\mathbf{q})(1-n_f(p_1^\ell))=2\omega_\mathbf{q}n_b(\omega_{\mathbf{q}}). \nonumber \\
\end{align}
Since each term in the $\mathbf{b}^\ell$, $\mathbf{b}_1^\ell$, $\mathbf{d}^\ell$ terms results in a convergent integral over $p^\ell_1$, the $\mathbf{v}_\ell^-\cdot\mathbf{q}$~s can be shifted out. Then, since $\sum_{\ell}\mathbf{v}^\pm_\ell=0$, the contribution from the $\mathbf{b}^\ell$, $\mathbf{b}_1^\ell$ and $\mathbf{d}^\ell$ terms vanishes. Then,
\begin{equation}
2(-i\omega+0^+)u(\omega,\mathbf{q},\omega_{\mathbf{q}})=-8\omega_\mathbf{q}\frac{g^2}{\pi v} u(\omega,\mathbf{q},\omega_{\mathbf{q}}).
\end{equation}
Since this has to hold for all values of $\omega$, we can only have $u=0$. Hence, the boson collision integral is trivially solved by the thermal Bose distribution and the bosons do not respond to the applied electric field in our approximation. It is also easily seen using the identity
\begin{equation}
n_f(x)(1-n_f(x-y))+n_b(y)(n_f(x)-n_f(x-y))=0,
\end{equation}
that the thermal Fermi distribution $n_f$ nullifies the fermion collision integral in the absence of an applied electric field if the thermal Bose distribution $n_b$ is used for the bosons, as it should.

\bibliography{sdw}

\end{document}